\documentclass{CSML}

\def\dOi{13(1:15)2017}
\lmcsheading%
{\dOi}
{1--36}
{}
{}
{Mar.~31, 2014}
{Mar.~24, 2017}
{}

\pdfoutput=1

\usepackage[hidelinks]{hyperref}
\usepackage{amssymb}
\usepackage{graphicx}
\usepackage{amsmath}
\usepackage{xspace}
\usepackage{mathtools}
\usepackage{multicol}

\usepackage{tikz}
\usetikzlibrary{arrows}
\usetikzlibrary{external}

\usepackage{enumitem}
\setenumerate{font=\normalfont,label=(\roman*)}

\newcommand{\prd}[1]{\Pi_{#1} }
\newcommand{\sm}[1]{\Sigma_{#1} }
\newcommand{\lam}[1]{\lambda #1 .}
\newcommand{\fa}[2]{\prd{#1:#2}}
\newcommand{\fasimple}[2]{\prd{#1 : #2}}
\newcommand{\fapairs}[3]{\fa {#1 , #2} #3}
\newcommand{\prdtypes}[1]{\prd{#1 : \UU}}
\newcommand{\jdeq}{\equiv}
\newcommand{\defeq}{\vcentcolon\equiv}
\newcommand{\eqvsym}{\simeq}
\newcommand{\eqv}[2]{\ensuremath{#1 \eqvsym #2}\xspace}

\newcommand{\eqvsymspace}{\enspace \eqvsym \enspace}
\newcommand{\eqvspace}[2]{\ensuremath{#1 \eqvsymspace #2}\xspace}
\newcommand{\idsym}{{=}}
\newcommand{\id}[3][]{\ensuremath{#2 =_{#1} #3}\xspace}
\newcommand{\idspace}[3][]{\ensuremath{#2 \, =_{#1} \, #3}\xspace}
\newcommand{\idtype}[3][]{\ensuremath{\mathsf{Id}_{#1}(#2,#3)}\xspace}

\newcommand{\refl}[1]{\ensuremath{\mathsf{refl}_{#1}}\xspace}

\newcommand{\ct}{%
  \mathchoice{\mathbin{\raisebox{0.5ex}{$\displaystyle\centerdot$}}}%
             {\mathbin{\raisebox{0.5ex}{$\centerdot$}}}%
             {\mathbin{\raisebox{0.25ex}{$\scriptstyle\,\centerdot\,$}}}%
             {\mathbin{\raisebox{0.1ex}{$\scriptscriptstyle\,\centerdot\,$}}}
}

\newcommand{\opp}[1]{\mathord{{#1}^{-1}}}
\newcommand{\trans}[2]{\ensuremath{{#1}_{*}\mathopen{}\left({#2}\right)\mathclose{}}\xspace}

\newcommand{\UU}{\ensuremath{\mathcal{U}}\xspace}
\newcommand{\N}{\ensuremath{\mathbb{N}}\xspace}
\newcommand{\emptyt}{\ensuremath{\mathbf{0}}\xspace}
\newcommand{\unit}{\ensuremath{\mathbf{1}}\xspace}

\newcommand{\bool}{\ensuremath{\mathbf{2}}\xspace}
\newcommand{\btrue}{{1_{\bool}}}
\newcommand{\bfalse}{{0_{\bool}}}

\newcommand{\iscontr}{\ensuremath{\operatorname{\mathsf{isContr}}}}
\newcommand{\isprop}{\ensuremath{\operatorname{\mathsf{isProp}}}}
\newcommand{\isset}{\ensuremath{\operatorname{\mathsf{isSet}}}}

\def\compare#1#2#3#4{\if#1#3\if#2#41\else0\fi\else0\fi}

\newcommand{\istype}[1]{
  \edef\a{\compare-2#1\empty\empty}
  \if\a1 \iscontr \else
  \edef\b{\compare-1#1\empty\empty}
  \if\b1 \isprop \else
  \edef\c{#1}
  \if0\c \isset \else
  \mathsf{is}\mbox{-}{#1}\mbox{-}\mathsf{type} \fi\fi\fi
}

\newcommand{\mapfunc}[1]{\ensuremath{\mathsf{ap}_{#1}}\xspace}

\newcommand{\dec}{\mathsf{decidable}}
\newcommand{\deceq}{\mathsf{isDiscrete}}

\newcommand{\fst}{\mathsf{fst}}
\newcommand{\snd}{\mathsf{snd}}
\newcommand{\inl}{\mathsf{inl}\xspace}
\newcommand{\inr}{\mathsf{inr}\xspace}

\newcommand{\LEM}[1]{\ensuremath{\mathsf{LEM}_{#1}}\xspace}
\newcommand{\UUpointed}{\UU_{\bullet}}
\newcommand{\image}{\mathsf{im}}

\newcommand{\trunc}[2]{\mathopen{}\left\Vert #2\right\Vert_{#1}\mathclose{}}
\newcommand{\ttrunc}[2]{\bigl\Vert #2\bigr\Vert_{#1}}

\newcommand{\tproj}[3][]{\mathopen{}\left|#3\right|_{#2}^{#1}\mathclose{}}

\newcommand{\brck}[1]{\trunc{}{#1}}
\newcommand{\bbrck}[1]{\ttrunc{}{#1}}

\newcommand{\bproj}[1]{\tproj{}{#1}}

\newcommand{\const}{\operatorname{\mathsf{wconst}}}
\newcommand{\coll}{\operatorname{\mathsf{constEndo}}}
\newcommand{\pathcoll}{\operatorname{\mathsf{pathConstEndo}}}
\newcommand{\stable}{\operatorname{\mathsf{stable}}}
\newcommand{\sep}{\operatorname{\mathsf{separated}}}
\newcommand{\hstable}{\operatorname{\mathsf{splitSup}}}
\newcommand{\istrans}{\operatorname{\mathsf{isTransitive}}}
\newcommand{\hsep}{\operatorname{\mathsf{hSeparated}}}
\newcommand{\fix}{\operatorname{\mathsf{fix}}}
\newcommand{\populated}[1]{\langle \! \langle #1 \rangle \! \rangle}

\newcommand{\h}{\operatorname{\mathsf{h_{tr}}}}
\newcommand{\elim}{\operatorname{\mathsf{rec_{tr}}}}
\newcommand{\depelim}{\operatorname{\mathsf{ind_{tr}}}}

\newcommand{\seg}{\operatorname{\mathsf{seg}}}
\newcommand{\I}{\mathbb I}
\newcommand{\myst}{\operatorname{\mathsf{myst}}}

\theoremstyle{plain}
\newtheorem{principle}[thm]{Principle}
\newtheorem{auxlem}[thm]{Auxiliary Lemma}

\begin{document}

\title[Notions of Anonymous Existence \mbox{in Martin-L\"of Type Theory}]
      {Notions of Anonymous Existence \mbox{in Martin-L\"of Type Theory}\rsuper*}

\author[N.~Kraus]{Nicolai Kraus\rsuper a}	
\address{{\lsuper{a,d}}University of Nottingham, School of Computer Science, Nottingham NG8 1BB, UK}	
\email{\{nicolai.kraus, thorsten.altenkirch\}@nottingham.ac.uk}  
\thanks{{\lsuper a}Supported by the EPSRC grant EP/M016994/1.} 

\author[M.~Escard\'{o}]{Mart\'{i}n H\"{o}tzel Escard\'{o}\rsuper b}	
\address{{\lsuper b}University of Birmingham, School of Computer Science, Birmingham B15 2TT, UK}	
\email{m.escardo@cs.bham.ac.uk}  

\author[T.~Coquand]{Thierry Coquand\rsuper c}	
\address{{\lsuper c}Chalmers University, Department of Computer Science and Engineering,
SE-412 96 G\"oteborg, Sweden}	
\email{thierry.coquand@cse.gu.se}  
\thanks{{\lsuper c}Supported by the ERC project 247219, and grants of The Ellentuck and The Simonyi Fund.}	

\author[T.~Altenkirch]{Thorsten Altenkirch\rsuper d}	
\address{\vspace{-18 pt}}	
\thanks{{\lsuper d}Supported by the EPSRC grants EP/G03298X/1 and EP/M016994/1 and by a grant of the Institute for Advanced Study, as well as by USAF, Airforce office for scientific research, award FA9550-16-1-0029.}	

\keywords{homotopy type theory, Hedberg's theorem, anonymous existence, weakly constant functions, factorization, truncation, squash types, bracket types, coherence conditions}
\subjclass{F.4.1 [Mathematical Logic and Formal Languages] - Lambda calculus and related systems} 
\titlecomment{{\lsuper *}Part of this article appeared as ``Generalizations of Hedberg's Theorem'' in the proceedings of Typed Lambda Calculus and Applications 2013}


\begin{abstract}
    As the groupoid model of Hofmann and Streicher proves, identity proofs in intensional Martin-L\"of type theory cannot generally be shown to be unique.
    Inspired by a theorem by Hedberg, we give some simple characterizations of types that do have unique identity proofs. 
    A key ingredient in these constructions are weakly constant endofunctions on identity types. 
    We study such endofunctions on arbitrary types and show that they always factor through a propositional type, the \emph{truncated} or \emph{squashed} domain.
    Such a factorization is impossible for weakly constant functions in general (a result by Shulman), but we present several non-trivial cases in which it can be done. 
    Based on these results, we define a new notion of anonymous existence in type theory and compare different forms of existence carefully. 
    In addition, we show possibly surprising consequences of the judgmental computation rule of the truncation, in particular in the context of homotopy type theory. 

    All the results have been formalized and verified in the dependently typed programming language Agda.
\end{abstract}

\maketitle

\section{Introduction}\label{sec1:introduction}

Although the identity type $\idtype a b$ is defined as an inductive type with only one single constructor $\refl{}$, 
it is a concept in Martin-L\"of type theory~\cite{Martin-Lof-1972}~\cite{Martin-Lof-1973}~\cite{Martin-Lof-1979} that is hard to get intuition for.
The reason is that it is, as a type family, parametrized twice over the same type, while the constructor only expects one argument:
$\refl a : \id a a$, where $\id a b$ is an alternative notation for $\idtype a b$. 
In fact, it is the simplest and most natural occurrence of this phenomenon. 

A result by Hofmann and Streicher~\cite{hofmannStreicher_groupoids} is that we can not prove $\refl a$ to be the only inhabitant of the type $\id a a$, that is, the principle of \emph{unique identity proofs} (UIP) is not derivable. 
Some time later, Hedberg~\cite{hedberg1998coherence} formulated a sufficient condition on a type to satisfy UIP, namely that its equality is decidable.

The core argument of the proof by Hofmann and Streicher is that types can be interpreted as groupoids, i.e.\ categories of which all morphisms are invertible.
Their conjecture that the construction could also be performed using higher groupoids was only made precise more that ten years later.
Awodey and Warren~\cite{awodeyWarren_HTmodelsOfIT} as well as, independently, Voevodsky~\cite{voevodsky_equivalenceAndUnivalence} explained that types can be regarded as, roughly speaking, topological spaces. Consequently, an exciting new direction of constructive formal mathematics attracted researchers from originally very separated areas of mathematics, and \emph{homotopy type theory}~\cite{HoTTbook} was born.

The current article is not only on homotopy type theory, but on Martin-L\"of type theory in general, even though we expect that the results are most interesting in the context of homotopy type theory. We start with Hedberg's Theorem~\cite{hedberg1998coherence} and describe multiple simple ways of strengthening it, one of them involving \emph{propositional truncation}~\cite{HoTTbook}, also known as \emph{bracket types}~\cite{awodeyBauer_bracketTypes} or \emph{squash types}~\cite{Con85}. 

Propositional truncation is a concept that provides a sequel to the \emph{Propositions-as-Types} paradigm~\cite{Howard80}. If we regard a type as the correspondent of a mathematical statement, a \emph{proposition}, and its inhabitants of proofs thereof, we have to notice that there is a slightly unsatisfactory aspect. A proof of a proposition in mathematics is usually not thought to contain any information apart from the fact that the proposition is true; however, a type can have any number of inhabitants, and therefore any number of witnesses of its truth. Hence it seems natural to regard only \emph{some} types as propositions, namely those which have at most one inhabitant. The notion of propositional truncation assigns to a type the proposition that this type is inhabited.
To make the connection clearer, these types are even called \emph{propositions}, or \emph{h-propositions}, in homotopy type theory. 
With this in mind, we want to be able to say that a type is inhabited without having to reveal an inhabitant explicitly. This is exactly what propositional truncation $\brck - : \UU \to \UU$ (where we write $\UU$ for the universe of types) makes possible. 
On the other hand, should $A$ have only one inhabitant up to the internal equality, this inhabitant can be constructed from an inhabitant of $\brck A$. This is a crucial difference between propositional truncation and double negation. 
We consider a weak version of $\brck -$ which does not have judgmental computation properties.

After discussing direct generalizations of Hedberg's Theorem, we attempt to transfer the results from the original setting, where they talk about equality types (of \emph{path spaces}), to arbitrary types. 
This leads to a broad discussion of \emph{weakly constant functions}: we say that $f: A \to B$ is \emph{weakly constant} if it maps any two elements of $A$ to equal elements of $B$. The attribute \emph{weakly} comes from the fact that we do not require these actual equality proofs to fulfil further conditions, and a weakly constant function does not necessarily appear to be constant in the topological models.
For exactly this reason, it is in general not possible to factor the function $f$ through $\brck A$; however, we can do it in certain special cases, and we analyze why. This has, for example, the consequence that the truncated sum of two proposition already has the universal property of their \emph{join}, which is defined as a \emph{higher inductive type} in homotopy type theory. 

Particularly interesting are weakly constant \emph{endofunctions}. We show that these can always be factored through the propositional truncation, based on the observation that the type of fixed points of such a function is a proposition. This allows us to define a new notion of existence which we call \emph{populatedness}. We say that $A$ is populated if any weakly constant endofunction on $A$ has a fixed point. This property is propositional and behaves very similar to $\brck A$, but we show that it is strictly weaker. On the other hand, it is strictly stronger than the double negation $\neg\neg A$, another notion of existence which, however, is often not useful as it generally only allows to prove negative statements. It is worth emphasizing that our populatedness is not a component that has to be \emph{added} to type theory, but a notion that can be \emph{defined} internally. We strongly suspect that this is not the case for even the weak version of propositional truncation, but we lack a formal proof.

It turns out to be interesting to consider the assumption that every type has a weakly constant endofunction.
The empty type has a trivial such endofunction, and so does a type of which we know an explicit inhabitant; however, from the assumption that a type has a weakly constant endofunction, we have no way of knowing in which case we are. 
In a minimalistic theory, we do not think that this assumption implies excluded middle. However, it implies that all equalities are decidable, i.e.\ a strong version of excluded middle holds for equalities.

Finally, we show that the judgmental computation rule of the propositional truncation, if it is assumed, does have some interesting consequences for the theory.
One of our observations is that we can construct a term $\myst_\N$ such that $\myst_\N(\bproj n)$ is judgmentally equal to $n$ for any natural number $n$, which shows that 
the projection map $\bproj - : \N \to \brck \N$ does not loose meta-theoretic information, in a certain sense.

Some parts of the Sections~\ref{sec3:hedbergs-theorem}, 
\ref{sec4:coll},
\ref{sec5:populatedness} and~\ref{sec6:taboos} of this article 
have been published in our previous conference paper~\cite{krausgeneralizations}.

\subsection*{Formalization} 

We have formalized~\cite{krausEscardoEtAll_existenceFormalisation} all of our results in the dependently typed programming language and proof assistant \emph{Agda}~\cite{Norell2007Towards}.
It is available in browser-viewable format and as plain source code on the first-named author's academic homepage.
All proofs type-check in Agda version 2.4.2.5.

As most of our results are internal statements in type theory, they can be formalized directly 
in a readable way, understandable even for readers who do not have any experience with the specific proof assistant or formalized proofs in general. 
We have tried our best and would like to encourage the reader to have a look at the accompanying formalization.

\subsection*{Contents}
In Section~\ref{sec2:preliminaries}, we specify the type theory that we work in, a standard version of Martin-L\"of type theory. We also state basic definitions, but we try to use standard notation and we hope that all notions are as intuitive as possible. We then revisit Hedberg's Theorem in Section~\ref{sec3:hedbergs-theorem} and formulate several generalizations.
Next, we move on to explore weakly constant functions between general types. 
We show that a weakly constant endofunction has a propositional type of fixed points and factors through $\brck -$ in Section~\ref{sec4:coll}. 
It is known that the factorization can not always be done for functions between different types, but we discuss some cases in which it is possible in Section~\ref{sec4c:factorizing}.
Section~\ref{sec5:populatedness} is devoted to \emph{populatedness}, a new definable notion of anonymous existence in type theory, based on our previous observations of weakly constant endofunctions. 
We examine the differences between inhabitance, populatedness, propositional truncation and double negation, all of which are notions of existence, carefully in Section~\ref{sec6:taboos}.
In particular, we show that if every type has a weakly constant endofunction, then all equalities are decidable.
Finally, Section~\ref{sec9:judgm-beta} discusses consequences of the judgmental computation rule of propositional truncation, and Section~\ref{sec10:open} presents a summary and questions which we do not know the answer to.

\section{Preliminaries}\label{sec2:preliminaries}

Our setting is a standard version of intensional Martin-L\"of type theory (MLTT) with type universes that have coproducts, dependent sums, dependent products and identity types. We give a very rough specification of these constructions below. For a rigorous treatment, we refer to our main reference \cite[Appendix A.1 or A.2]{HoTTbook}. 
We use standard notation whenever it is available. If it improves the readability, we allow ourselves to implicitely uncurry functions and write $f(x,y)$ instead of $f(x)(y)$ or $f\,x\,y$.

\emph{Type Universes.}
MLTT usually comes equipped with a hierarchy $\UU_0, \UU_1, \UU_2, \ldots$ of universes, where $\UU_{n+1}$ is the type of $\UU_n$. 
With very few exceptions, we only need one universe $\UU$ and therefore omit the index. $\UU$ can be understood as a generic universe or, for simplicity, as the lowest universe $\UU_0$.
If we say that $X$ is a type, we mean $X:\UU$, possibly in some context.

\emph{Coproducts.}
If $X$ and $Y$ are types, then so is $X+Y$. If we have $x:X$ or $y:Y$, we get $\inl\, x : X+Y$ or $\inr\, y : X+Y$, respectively. To prove a statement for all elements in $X+Y$, it is enough to consider those that are of one of these two forms.

\emph{Dependent Pairs.}
If $X$ is a type and $Y : X \to \UU$ a family of types, indexed over $X$, then $\sm{X}Y$ is the corresponding \emph{dependent pair type}, sometimes called a \emph{dependent sum} or just \emph{$\Sigma$-type}. For $x:X$ and $y:Y(x)$, we have $(x,y): \sm{X}Y$, and to eliminate out of $\sm{X}Y$, it is enough to consider elements of this form. 
We prefer to write $\sm{x:X}Y(x)$ instead of $\sm{X}Y$, hoping to increase readability.
Instead of $\sm{x_1:X}\sm{x_2:X}Y(x_1,x_2)$, we write $\sm{x_1,x_2:X}Y(x_1,x_2)$.
In the special case that $Y$ does not depend on $X$, it is standard to write $X \times Y$. 

\emph{Dependent Functions.}
Given $X: \UU$ and $Y:X \to \UU$ as before, we have the type $\prd{X}Y$, called the \emph{dependent functions type} or $\Pi$-type. It is sometimes also referred to as the \emph{dependent product type}, although that notion can be confusing as it would fit for $\Sigma$-types as well.
If, for any given $x:X$, the term $t$ is an element in $Y(x)$, we have $\lam{x}t : \prd{X}Y$. 
Similarly to $\prd{X}Y$, we write $\prd{x:X}Y(x)$, and, if $Y$ does not depend on $X$, we write $X \to Y$.
Instead of $\prd{x_1:X}\prd{x_2:X} Y(x_1,x_2)$, we write $\prd{x_1,x_2:X}Y(x_1,x_2)$.

\emph{Identity Types.}
Given a type $X$ with elements $x,y:X$, we have the \emph{identity type} or the type of \emph{equalities}, written $\id [X] {x} {y}$. 
An inhabitant $p : \id[X] x y$ is thus called an \emph{equality}, an \emph{equality proof}, or, having the interpretation of a type as a space in mind, a \emph{path} from $x$ to $y$. Similarly, $\id[X] x y$ is called a \emph{path space}. 
In the past, $p$ often used to be called a \emph{propositional} equality. We avoid this terminology and reserve the word ``propositional'' for types with at most one element, as explained in the introduction and in Definition~\ref{def:generalnotions}.

The only introduction rule for the identity types is that, for any $x:X$, there is $\refl x : \id [X] x x$. The elimination rule (called \emph{J}) says that, if $P : (\sm{x,y:X} \id[X] x y) \to \UU$ is a type family, it suffices to construct an inhabitant of $\prd{x:X}P(x,x,\refl x)$ in order to get an element of $P(p)$ for any $p : \sm{x,y:X} \id[X] x y$. 
We do explicitly not assume other elimination rules such as \emph{Streicher's K} 
or \emph{uniqueness of identity proofs (UIP)}~\cite{Streicher93}. If the common type of $x,y$ can be inferred or is unimportant, we write $\id x y$ instead of $\id [X] x y$.

In contrast to the identity type, \emph{definitional} (also called \emph{judgmental}) equality is a meta-level concept. 
It refers to two terms, rather than two (hypothetical) elements, with the same $\beta$ (and, sometimes, $\eta$ in a restricted sense) normal form. 
Recently, it has become standard to use the symbol $\jdeq$ for judgmental equality in order to use $\idsym$ solely for the type of equalities~\cite{HoTTbook}. 
Note that the introduction rule of the latter says precisely that we have a canonical equality proof for any two judgmentally equal terms, viewed as elements of some type. For definitions, we use the notation $\defeq$.

Applying the eliminator \emph{J} is also referred to as \emph{path induction}~\cite{HoTTbook}. A variant of \emph{J} that is sometimes more useful is due to Paulin-Mohring~\cite{Moh93}: given a point $x:X$ and a type family $P : (\sm{y:X} \id[X] x y) \to \UU$, it is enough to construct an inhabitant of $P(x,\refl x)$ in order to construct an inhabitant of $P(y,q)$ for any pair $(y,q)$. This elimination principle, called \emph{based path induction}, is equivalent to \emph{J}.

As a basic example, we show that equality proofs satisfy the \emph{groupoid laws}~\cite{hofmannStreicher_groupoids}, where reflexivity plays the role of identity morphisms. 
If we have $p : \id [X] x y$ and $q : \id [X] y z$, we can construct a path $p \ct q : \id [X] x z$ (the \emph{composition} of $p$ and $q$): by based path induction, it is enough to do this under the assumption that $(z,q) : \sm{z:X} \id [X] y z$ is $(y, \refl y)$. But in that case, the composition $p \ct q$ is given by $p$. Similarly, for $p : \id [X] x y$, there is $\opp p : \id [X] y x$. It is easy to see (again by path induction) that the types $\id [X] {p \ct {\refl y}} p$ and $\id [X] {{\refl x} \ct p} p$ as well as $\id [X] {p \ct \opp p} {\refl x}$ are inhabited, and similarly, so are all the other types that are required to give a type the structure of a groupoid. 

An important special case of the eliminator \emph{J} is \emph{substitution} or \emph{transportation}: 
if $P : X \to \UU$ is a family of types and $x,y : X$ are two elements (or points) that are equal by $p : \id [X] x y$, 
then an element of $e : P(x)$ can be \emph{transported along the path $p$} to get an element of $P(y)$, written
\begin{equation}
\trans{p}{e} : P(y). 
\end{equation}
Another useful function, similarly easily derived from \emph{J}, is the following: if $f : X \to Y$ is a function and $p : \id [X] {x}{y}$ a path, we get an inhabitant of $\id{f(x)}{f(y)}$ in $Y$,
\begin{equation}
\mapfunc{f}{p} : \id {f(x)} {f(y)}.
\end{equation}
Note that we omit the arguments $x$ and $y$ in the notation of $\mapfunc f$.

Identity types also enable us to talk about isomorphism, or (better) \emph{equivalence}, 
of types. 
We say that $X$ and $Y$ are equivalent, written $X \simeq Y$, if there are functions in both directions which are the inverses of each other,
\begin{align}
 &f : X \to Y \label{eq:f-part-of-equivalence} \\
 &g : Y \to X \label{eq:g-part-of-equivalence} \\
 &p : \fa x X \id[X]{g(f(x))}{x} \\
 &q : \fa y Y \id[Y]{f(g(y))}{y}.
\end{align}
Technically, $(f,g,p,q)$ only constitute what is usually called a \emph{type isomorphism}, but from any such isomorphism, an equivalence (in the sense of homotopy type theory) can be constructed; and the only difference is that an equivalence requires a certain coherence between the components $p$ and $q$, which will not be important for us. In this sense, we do not distinguish between isomorphims and equivalences, and only choose the latter terminology on principle.
For details, we refer to~\cite[Chapter 4]{HoTTbook}. 
We call types \emph{logically equivalent}, written $X \Leftrightarrow Y$, if there are functions in both directions (that is, we only have the components~\eqref{eq:f-part-of-equivalence} and~\eqref{eq:g-part-of-equivalence}). We write $X \Leftrightarrow Y \Leftrightarrow Z$ if $X,Y,Z$ are pairwise logically equivalent, and $X \Rightarrow Y \Rightarrow Z$ as a shorthand notation for $(X \to Y) \times (Y \to Z)$.

Equivalent types share all internalizable properties. In fact, Voevodsky's univalence axiom (e.g.~\cite{HoTTbook},~\cite{voevodsky_equivalenceAndUnivalence}) has the consequence that equivalent types are equal. 
For the biggest part of our article, we do not need to assume the univalence axiom; however, it will play some role in Section~\ref{sec9:judgm-beta}.

We sometimes use other additional principles (namely function extensionality and propositional truncation, as introduced later). 
However, we treat them as assumptions rather than parts of the core theory and state clearly in which cases they are used.

In order to support the presentation from the next section on, we define a couple of notions.
Our hope is that all of these are as intuitive as possible, if not already known. 
The only notion that is possibly ambiguous is \emph{weak constancy}, 
meaning that a function maps any pair of possible arguments to equal values.

\begin{defi} \label{def:generalnotions}
 We say that a type $X$ is \emph{propositional}, or is a \emph{proposition},
if all its inhabitants are equal: 
 \begin{equation}
  {\istype {-1} X} \defeq {\fa{x,y}{X} \id x y}.
 \end{equation}
 It is a well-known fact that the path spaces of a propositional type are not only inhabited but also propositional themselves.
 This stronger property is called \emph{contractible},
 \begin{equation}
  {\istype {-2} X} \defeq X \times \istype {-1} X.
 \end{equation}
 It is easy to see that any contractible type is equivalent to the unit type.
 An important well-known lemma is that types are contractible if they are represented as \emph{singletons}, sometimes called \emph{path-to}/\emph{path-from} types: 
 for any $a_0 : A$, the type
 \begin{equation}
  \sm{a:A}\id{a_0}{a}
 \end{equation}
is contractible, as any inhabitant is by based path induction easily seen to be equal to $(a_0, \refl {a_0})$.
 
Further, $X$ satisfies \emph{UIP}, or is a \emph{set}, if its path spaces are all propositional:
\begin{equation}
 \istype {0} X \defeq \fa{x,y}{X} \istype {-1} (\id x y).
\end{equation}

$X$ is \emph{decidable} if it is either inhabited or empty,
 \begin{equation}
  \dec X \defeq X + \neg X.
 \end{equation}
We therefore say that $X$ has \emph{decidable equality} if the equality type of any two inhabitants of $X$ is decidable.
Based on the terminology in~\cite{mines_constAlgebra}, we also call a type with decidable equality \emph{discrete}:
\begin{equation}
 \deceq X \defeq \fa{x,y}{X} \dec (\id x y).
\end{equation}

A function (synonymously, map) $f : X \to Z$ is \emph{weakly constant}, or \emph{1-constant}, if it maps any two elements to the same inhabitant of $Y$:
\begin{equation}
 \const f \defeq \fa{x,y}{X} \id{f(x)}{f(y)}.
\end{equation}
As weak (or 1-) constancy is the only notion of constancy that we consider in this article (if we ignore factorizability through $\brck -$), we call such a function $f$ just \emph{constant} for simplicity. 
However, note that this notion is indeed very weak as soon as we consider functions into types that are not sets, as we will see later.
It will be interesting to consider the type of constant endomaps on a given type:
\begin{equation}
 \coll X \defeq \sm{f: X \to X} \const f.
\end{equation}
Finally, we may say that $X$ has constant endomaps on all path spaces: 
\begin{equation}
 \pathcoll X \defeq \fa{x,y}{X} \coll {(\id x y)}.
\end{equation}
\end{defi}

For some statements, but only if clearly indicated, we use \emph{function extensionality}. This principle says that two functions $f, g$ of the same type $\prd X Y$ 
are equal as soon as they are pointwise equal:
\begin{equation} \label{eq:naive-funext}
 \left(\fasimple{x} X \id{f(x)}{g(x)}\right) \to \id f g.
\end{equation}
An important equivalent formulation due to Voevodsky~\cite{voe_coqLib} is that the type of propositions is closed under $\Pi$; more precisely, 
\begin{equation} \label{eq:voe-funext}
 \left(\fasimple{x} X \istype {-1} \left(Y \, x\right)\right) \, \to \, \istype {-1} \left(\prd{X}Y\right).
\end{equation}
In the case of non-dependent functions, this means that $X \to Y$ is propositional as soon as $Y$ is.

A principle that we do not assume, but which will appear in some of our discussions, is the \emph{law of excluded middle} in the form for propositions and in the form for general types \cite[Chapter 3.4]{HoTTbook}. 
In the first form, it says that every proposition is decidable, while the second says the the same without the restriction to propositions.
\begin{align}
 &\LEM {} \defeq \prdtypes {P} (\istype {-1} P) \to P + \neg P \\ 
 &\LEM {\infty} \defeq \prdtypes {X} X + \neg X.
\end{align}
Note that $\LEM \infty$ can be considered the natural formulation under the \emph{Propositions-as-Types} view. 
However, the view that we adapt in this work (as in homotopy type theory) is the one that only type-theoretical propositions in the sense of Definition~\ref{def:generalnotions} really represent mathematical propositions; general types carry more structure. 
In particular, $\LEM \infty$ includes a very strong form of choice which is inconsistent with the univalence axiom of homotopy type theory. 
Therefore, we consider $\LEM {}$ the ``correct'' formulation in our work.

We do not explicitly use this fact, but it may be helpful to note that, assuming function extensionality, all of the above definitions that are called ``is\ldots'' ($\istype{-1} X$, $\istype{-2} X$, $\istype{0} X$, $\deceq X$) are propositional in the sense of Definition~\ref{def:generalnotions}. For $\istype{-1} X$, $\istype{-2} X$, $\istype{0} X$, this is proved in~\cite[Theorem 7.1.10]{HoTTbook}, and for $\deceq X$, this is a consequence of Hedberg's Theorem that we discuss in Section~\ref{sec3:hedbergs-theorem}.
It will also follow that $\pathcoll X$ is propositional.
The statements of $\LEM{}$ and \eqref{eq:voe-funext} are propositional as well, while $\const f$, $\coll X$, \eqref{eq:naive-funext}, and $\LEM\infty$ are in general not propositional.

\section{Hedberg's Theorem}\label{sec3:hedbergs-theorem}

Before discussing possible generalizations, we want to state Hedberg's Theorem.
\begin{thm}[{Hedberg~\cite{hedberg1998coherence}}] \label{thm:hedberg}
 Every discrete type satisfies UIP,
\begin{equation}
  \deceq X \to \istype 0 X.
 \end{equation}
\end{thm}
We briefly give Hedberg's original proof, consisting of two steps.
\begin{lem} \label{lem:discr2pathcoll}
 If a type has decidable equality, its path spaces have constant endofunctions:
 \begin{equation}
  \deceq X \to \pathcoll X.
 \end{equation}
\end{lem}
\proof
 Given inhabitants $x$ and $y$ of $X$, we get by assumption either an inhabitant of $\id x y$ or an inhabitant of $\neg (\id x y)$.
 In the first case, we construct the required constant function $(\id x y) \to (\id x y)$ by mapping everything to this given path. 
 In the second case, we have a proof of $\neg (\id x y)$, and the canonical function is constant automatically.
\qed
\begin{lem}\label{lem:pathcoll2set}
 If the path spaces of a type have constant endomaps, the type satisfies UIP:
 \begin{equation}
  \pathcoll X \to \istype 0 X.
 \end{equation}
\end{lem}
\proof
 Assume $f$ is a parametrized constant endofunction on the path spaces,
 meaning that, for any $x,y:X$, we have a constant function $f_{x,y} : \id x y \to \id x y$.
 Let $p$ be a path from $x$ to $y$. We claim that
\begin{equation} \label{eq:pathcoll-set-proof}
 \id {p}  {{\opp{(f_{x,x}(\refl x))}} \ct {f_{x,y}(p)}}.
\end{equation}
By path induction, we only have to give a proof if the triple $(x,y,p)$ is in fact $(x, x, \refl x)$, in which case~\eqref{eq:pathcoll-set-proof} is one of the groupoid laws that equality satisfies. 
Using the fact $f$ is constant on every path space, the right-hand side of the above equality is independent of $p$, and in particular, equal to any other path of the same type.
\qed
Hedberg's proof~\cite{hedberg1998coherence} is just the concatenation of the two lemmata. A slightly more direct proof can be found in the HoTT Coq repository~\cite{hott_coqLib} and in a post by the first named author on the HoTT blog~\cite{nicolai:blog}.

Let us analyse the ingredients of the original proof.
Lemma~\ref{lem:discr2pathcoll} uses the rather strong assumption of decidable equality.
In contrast, the assumption of Lemma~\ref{lem:pathcoll2set} is logically equivalent to its conclusion, so that there is no space for a strengthening. We include a proof of this simple claim in Theorem~\ref{tfae} below and concentrate on weakening the assumption of Lemma~\ref{lem:pathcoll2set}. 
Let us first introduce the notions of \emph{stability} and \emph{separatedness}.
\begin{defi}
For a type $X$, define
\begin{align}
 &\stable X  \defeq \neg\neg X \to X, \\
 &\sep X  \defeq \fa{x,y}{X} \stable (\id x y).
\end{align}
\end{defi}
We can see $\stable X$ as a classical condition, similar to $\dec X \jdeq X + \neg X$, but strictly weaker. 
Indeed, we get a first strengthening of Hedberg's Theorem as follows:

\begin{lem}[{{\cite[Corollary 7.2.3]{HoTTbook}}}] \label{lemmaSepExtUip}
 Assuming function extensionality, any separated type is a set, 
 \begin{equation}
  \sep X \to \istype 0 X.
 \end{equation}
\end{lem}
\proof
 There is, for any $x, y : X$, a canonical map ${(\id x y) \to \neg \neg (\id x y)}$. Composing this map with the proof that $X$ is separated yields an endofunction on the path spaces. With function extensionality, the first map has a propositional codomain, implying that the endofunction is constant and thereby fulfilling the requirements of Lemma~\ref{lem:pathcoll2set}.
\qed

We remark that full function extensionality is actually not needed here. Instead, a weaker version that only works with the empty type is sufficient.
Similar statements hold true for all further applications of extensionality in this paper.

In a constructive setting, the question how to express that ``there
exists something'' in a type $X$ is very subtle. One possibility is to
ask for an inhabitant of $X$, but in many cases, this is too strong to be fulfilled. 
A second possibility, which corresponds to our
above definition of \emph{separated}, is to ask for a proof of $\neg
\neg X$. Then again, this is very weak, and often too weak, as one can
in general only prove negative statements from double-negated
assumptions.

This fact has inspired the introduction of \emph{squash types} (Constable~\cite{Con85}), and similar, \emph{bracket types} (Awodey and Bauer~\cite{awodeyBauer_bracketTypes}). These lie in between of the two extremes mentioned above. In our intensional setting, we talk of \emph{propositional truncations}, or \emph{$-1$-truncations} \cite[Chapter 3.7]{HoTTbook}. 
For any type $X$, we postulate that there is a type $\brck X$ that is a \emph{proposition}, representing the statement that $X$ is inhabited. 
The rules are that if we have a proof of $X$, we can, of course, get a proof of $\brck{X}$, and from $\brck X$, we can conclude the same statements as we can conclude from $X$, but only if the actual representative of $X$ does not matter:
\begin{defi}\label{def:htruncation}
 We say that a type theory has \emph{weak propositional truncations} if, for every type $X$, we have a type $\brck X : \UU$ which satisfies the following properties:
 \begin{enumerate}
  \item $\bproj{-} : X \to \brck X$  \label{item:trunc1}
  \item $\h : \istype {-1} (\brck{X})$ \label{item:trunc2}
  \item $\elim : \fa P \UU \istype {-1} P \to (X \to P) \to \brck X \to P.$\label{item:trunc3}
 \end{enumerate}
\end{defi}
 
\noindent Note that this amounts to saying that the operator $\brck{-}$ is left adjoint to the inclusion of the subcategory of propositions into the category of all types. Therefore, it can be seen as the \emph{propositional reflection}. 
 For $x,y:\brck X$, we will write $\h_{x,y}$ for the proof of $\id[\brck X] x y$ that we get from $\h$.

In contrast to other sources~\cite{HoTTbook} we do \emph{not} assume the judgmental $\beta$-rule 
 \begin{equation} 
  \elim (P, h, f, \bproj x) \; \jdeq_\beta \; f(x) \label{eq:jdgm-beta-1}
 \end{equation}
as it is simply not necessary for our results and we do not want to make the theory stronger than required. This is the reason why we use the attribute \emph{weak}.
We do think that~\eqref{eq:jdgm-beta-1} is often useful, but we also think it is interesting to make clear in which sense~\eqref{eq:jdgm-beta-1} makes the theory actually stronger, rather than more convenient. We will discuss this in Section~\ref{sec9:judgm-beta}.
A practical advantage of not assuming~\eqref{eq:jdgm-beta-1} is that the truncation can be implemented in existing proof assistants more easily.
Of course, the $\beta$-rule holds propositionally as both sides of the equation inhabit the same proposition.

Adopting the terminology of \cite[Chapter 3.10]{HoTTbook}, we say that $X$ is \emph{merely} inhabited if $\brck X$ is inhabited. We may also say that $X$ \emph{merely} holds. However, we try to always be precise by giving the formal type expression to support the informal statement.

The \emph{non-dependent eliminator} (or \emph{recursion principle}, see~\cite[Chapter 5.1]{HoTTbook}) $\elim$ lets us construct the \emph{dependent} one (the \emph{induction principle}):
 \begin{lem} [{{see~\cite[Exercise 3.17]{HoTTbook}}}] \label{lem:ind-from-rec}
 The propositional truncation admits the following induction principle:
 Given a type $X$, a family $P : \brck X \to \UU$ with $h : \prd{z: \brck X} \istype {-1} (P(z))$, a term $f : \prd{x:X} P(\bproj x)$ gives rise to an inhabitant of $\prd{z: \brck X} P(z)$.  
\end{lem}
\proof
We have a map $j : X \to \sm{z:\brck X}P(z)$ by $\lam x (\bproj x, f(x))$.
Observe that the codomain of $j$ is a proposition, combining the fact that $\brck X$ is one with $h$. 
Therefore, we get $\brck X \to \sm{z:\brck X}P(z)$, and this is sufficient, using that $\id[\brck X]yz$ for any $y,z:\brck X$.
\qed
In analogy to the notation $\elim$, we may write $\depelim$ for the term witnessing this induction principle. 
However, most of our further developments will not require the induction principle and will be proved with $\elim$.

Note that $\brck{-}$ is functorial in the sense that any function $f : X \to Y$ gives rise to a function $\brck f : \brck X \to \brck Y$, although the proof of $\id{\brck {g \circ f}}{\brck g \circ \brck f}$ requires function extensionality.
It is easy to see that $\brck{-}$ is a modality (an idempotent monad) in the sense of~\cite[Chapter 7.7]{HoTTbook}. In particular, we have $\eqv{\brck {\brck X}}{\brck X}$.

It is well-known that there is a type expression which is logically equivalent to the propositional truncation:
\begin{thm}\label{hinhabitedLargeSmall}
 For any given $X : \UU$, we have the logical equivalence
  \begin{equation} \label{eq:hinhabitedLargeSmall}
   \brck{X} \; \Leftrightarrow \; \prdtypes{P} \istype {-1} P \to (X \to P) \to P.
  \end{equation}
\end{thm}
Under the assumption of function extensionality, the expression on the right-hand side of~\eqref{eq:hinhabitedLargeSmall} is propositional, and the logical equivalence ($\Leftrightarrow$) is thus an actual equivalence.
A potential problem with this expression is that it does not live in the universe $\UU$. This size issue is the only thing that keeps us from using it as the definition for $\brck X$. All other properties of the above Definition~\ref{def:htruncation} are satisfied, at least under the assumption of function extensionality. Voevodsky~\cite{voe_coqLib} suggests \emph{resizing rules} to resolve the issue. 
\proof[Proof of Theorem~\ref{hinhabitedLargeSmall}]
 The direction ``$\rightarrow$'' of the statement is no more than a rearrangement of the assumptions of property~\ref{item:trunc3} in the definition of $\brck X$. For the other direction, we only need to instantiate $P$ with $\brck X$ and observe that the properties~\ref{item:trunc1} and~\ref{item:trunc2} are exactly what is needed.
\qed

With this definition at hand, we can provide an even stronger variant of Hedberg's Theorem. 
Completely analogously to the notions of stability and separatedness, we define what is means to say that a type has \emph{split support} and is \emph{h-separated}:
\begin{defi} \label{def:hsep}
For a type $X$, define
\begin{align}
 &\hstable X  \defeq \brck X \to X, \\
 &\hsep X  \defeq \fa{x,y}{X} \hstable (\id x y).
\end{align}
\end{defi}

We observe that $\hsep X$ is a weaker condition than $\sep X$. Not only can we conclude $\istype 0 X$ from $\hsep X$, but the converse holds as well. 
In the following theorem, we also include the simple fact that having constant endomaps on path spaces is equivalent to these statements. 
\begin{thm}\label{tfae}
 For a type $X$ in MLTT with propositional truncation, the following are equivalent:
 \begin{enumerate}
  \item $X$ is a set \label{item:tfae-1} 
  \item $X$ has constant endomaps on its path spaces \label{item:tfae-2} 
  \item $X$ is h-separated. \label{item:tfae-3}
 \end{enumerate}
 Further, each of the three types is propositional. 
\end{thm}

\proof
 We first show the logical equivalence of the three types.
 ``\ref{item:tfae-2} $\Rightarrow $~\ref{item:tfae-1}'' is simply Lemma~\ref{lem:pathcoll2set}. 
 ``\ref{item:tfae-1} $\Rightarrow $\ref{item:tfae-3}'' uses simply the the definition of the propositional truncation: given $x, y : X$, the fact that $X$ is a set tells us exactly that $\id x y$ is propositional, implying that we have a map $\brck{\id x y} \to (\id x y)$. 
 Concerning ``\ref{item:tfae-3} $\Rightarrow$~\ref{item:tfae-2}'', it is enough to observe that the composition of $\bproj{-} : (\id x y) \to \brck {\id x y}$ and the map $\brck{\id x y} \to (\id x y)$, provided by the fact that $X$ is h-separated, is a parametrized constant endofunction.
 
 \ref{item:tfae-1} is known to be a proposition. If~\ref{item:tfae-2} or~\ref{item:tfae-3} are inhabited, then $X$ is a set, implying that~\ref{item:tfae-2} and~\ref{item:tfae-3} are propositions.
\qed

We observe that using propositional truncation in some cases makes it unnecessary to appeal to functional extensionality.
In Lemma~\ref{lemmaSepExtUip}, we have given a proof for the simple statement that separated types are sets in the context of function extensionality. 
Let us now drop function extensionality and assume instead that propositional truncation is available. Every separated type is h-separated --- more generally, we have
 \begin{equation} \label{eq:negnegXsep}
  (\neg\neg X \to X) \to (\brck{X} \to X)
 \end{equation}
 for a type $X$ --- and every h-separated space is a set. Notice
 that $\neg X \to \neg \brck X$ and thus also $\brck X \to \neg \neg X$ and~\eqref{eq:negnegXsep} do not require function extensionality.
 Therefore, the mere availability of propositional truncation suffices to
 solve a gap that function extensionality would usually fill.
 In Section~\ref{subsec:beta-funext} below, we will see that propositional truncation with the judgmental $\beta$-rule \eqref{eq:jdgm-beta-1} makes it possible to derive function extensionality.

A variant of Theorem~\ref{tfae}, more precisely of the direction ``\ref{item:tfae-3} $\Rightarrow $\ref{item:tfae-1}'', can be formulated without propositional truncation.
We say that a \emph{reflexive propositionally-valued relation} on $X$ is an $R : X \times X \to \UU$ such that $R(x,y)$ is always propositional and $R(x,x)$ always contractible.
If $R$ implies identity, that is $\prd{x,y:X} R(x,y) \to \id x y$, then $X$ is a set. This is a statement given in the standard text book on homotopy type theory~\cite[Theorem 7.2.2]{HoTTbook} and is sometimes called ``Rijke's Theorem''.

To conclude this part of the article, we want to mention that there is a slightly stronger version of Hedberg's Theorem which applies to types where equality might only be decidable \emph{locally}. 
In fact, nearly everything we stated or proved can be done locally, and thus made stronger.
In the proof of Lemma~\ref{lem:discr2pathcoll}, we have not made use of the fact that we were dealing with path spaces at all: any decidable type trivially has a constant endofunction.
Concerning Lemma~\ref{lem:pathcoll2set}, we observe:
\begin{lem}[Local form of Lemma~\ref{lem:pathcoll2set}]
 A type $X$ that locally has constant endomaps on path spaces does locally satisfy UIP. 
 That means, for any $x_0 : X$, we have
 \begin{equation}
  (\fasimple y X \coll(\id {x_0} y)) \to \fasimple y X \istype {-1} (\id {x_0} y).
 \end{equation}
\end{lem}
\proof
The proof is identical to the one of Lemma~\ref{lem:pathcoll2set}, with the only difference that we need to apply based path induction instead of path induction.
\qed
This enables us to prove the local variant of Hedberg's Theorem:
\begin{thm}[{\cite{DBLP:journals/aml/Palmgren12},\cite{nicolai:blog}}; Local form of Theorem~\ref{thm:hedberg}]
 A locally discrete type $X$ is locally a set, i.e.\ for any $x_0 : X$,
\begin{equation}
 (\fasimple {y} X \dec (\id {x_0} y)) \to \fasimple y X \istype {-1} (\id {x_0} y).
\end{equation}
\qed
\end{thm}
In the same simple way, we immediately get that the assumption of local separatedness is sufficient.
\begin{lem}[Local form of Lemma~\ref{lemmaSepExtUip}]
 Under the assumption of function extensionality, a locally separated type locally is a set, i.e.\ for any $x_0 : X$,
 \begin{equation}
  (\fasimple y X \stable (\id {x_0} y)) \to \fasimple y X \istype {-1} (\id {x_0} y).
 \end{equation} 
\qed
\end{lem}
Similarly, the local forms of the characterizations of Theorem~\ref{tfae} are still equivalent.
\begin{thm}[Local form of Theorem~\ref{tfae}]
 For a type $X$ in MLTT with propositional truncation with a point $x_0:X$, the following are equivalent:
 \begin{enumerate}
  \item for all $y:X$, the type $\id {x_0} y$ is propositional \label{item:tfae-local-1} 
  \item for all $y:X$, the type $\id {x_0} y$ has a constant endomap \label{item:tfae-local-2} 
  \item for all $y:X$, the type $\id {x_0} y$ has split support.  \label{item:tfae-local-3} \qed 
 \end{enumerate}
\end{thm}

Note that most of our arguments can be generalized to higher truncation levels \cite[Chapter 7]{HoTTbook} in a reasonable and straightforward way. Details can be found in the first-named author's PhD thesis~\cite{nicolai:thesis}.

\section{Split Support from Constant Endofunctions}\label{sec4:coll}

If we unfold the definitions in the statements of Theorem~\ref{tfae}, they all involve the path spaces over some type $X$:
 \begin{enumerate}
  \item $\fa {x,y} X \istype {-1} (\id x y)$ \label{item:tfae-again-1}
  \item $\fa {x,y}{X} \coll {(\id x y)}$ \label{item:tfae-again-2}
  \item $\fa {x,y}{X} \hstable (\id x y)$. \label{item:tfae-again-3}
 \end{enumerate}
We have proved that these statements are logically equivalent. 
It is a natural question to ask whether this is true for types that are not necessarily path spaces.
The possibilities that path spaces offer are very
powerful and we have used them heavily. Indeed, if we formulate the
above properties for an arbitrary type $A$ instead of path types,
 \begin{enumerate}
  \item $\istype {-1} A$  \label{item:tfae-aga-1-prime}
  \item $\coll {A}$ \label{item:tfae-aga-2-prime}
  \item $\hstable A$,  \label{item:tfae-aga-3-prime}
 \end{enumerate}
we notice immediately that~\ref{item:tfae-aga-1-prime} is significantly and strictly stronger than the other two properties.~\ref{item:tfae-aga-1-prime} says that $A$ has at most one inhabitant, 
\ref{item:tfae-aga-2-prime} says that there is a constant endofunction on $A$, and~\ref{item:tfae-aga-3-prime}
gives us a possibility to get an explicit inhabitant of A from the proposition that A has an anonymous inhabitant.
A propositional type has the other two properties trivially, while the converse is not true. In fact, as soon as we know an inhabitant $a : A$, we can very easily construct proofs of~\ref{item:tfae-aga-2-prime} and~\ref{item:tfae-aga-3-prime}, while it does not help at all with~\ref{item:tfae-aga-1-prime}. 

The implication~\ref{item:tfae-aga-3-prime} $\Rightarrow$~\ref{item:tfae-aga-2-prime} is also simple: if we have $h : \brck{A} \to A$, the composition $h \circ \bproj{-} : A \to A$ is constant, as for any $a, b : A$, we have $\id{\bproj a}{\bproj b}$ and therefore $\id{h(\bproj a)}{h(\bproj b)}$.

In summary, we have~\ref{item:tfae-aga-1-prime} $\Rightarrow$~\ref{item:tfae-aga-3-prime} $\Rightarrow$~\ref{item:tfae-aga-2-prime} and we
know that the first implication cannot be reversed. What is less clear
is the reversibility of the second implication: If we have a constant
endofunction on $A$, can we get a map $\brck A \to A$? Put
differently, what does it take to get out of $\brck A$? Of course,
a proof that $A$ has split support is fine for that, but does a constant
endomap on $A$ also suffice?  Surprisingly, the answer is
positive, and there are interesting applications
(Section~\ref{sec5:populatedness}).
The main ingredient of our proof, and of much of the rest of the paper, is the following crucial lemma about fixed points:

\begin{lem}[Fixed Point Lemma]\label{fixedpoint}
 Given a constant endomap $f$ on a type~$X$, the type of its fixed points is propositional, where this type is defined by
 \begin{equation}
  \fix f \defeq \sm{x:X}{\id{x}{f(x)}}.
 \end{equation}
\end{lem}
Before we can give the proof, we first need to formulate two observations. Both of them are simple on their own, but important insights for the Fixed Point Lemma. Let $X$ and $Y$ be two types. 
\begin{auxlem}[{\cite[Theorem 2.11.3]{HoTTbook}}]\label{one}
Assume $h, k: X \to Y$ are two functions and $t : \id x y$ as well as $p : \id{h(x)}{k(x)}$ are paths. 
Then, transporting along $t$ into $p$ can be expressed as a composition of paths:
\begin{equation}
 \id{\trans t p} {\opp{(\mapfunc h t)} \ct p \ct \mapfunc k t}. 
\end{equation}
\end{auxlem}
\proof
This is immediate by path induction on $t$. 
\qed
Even if the latter proof is trivial, the statement is essential. In the proof of Lemma~\ref{fixedpoint}, we need a special case where $x$ and $y$ are the same. However, this special version cannot be proved directly. We consider the second observation a key insight for the Fixed Point Lemma:
\begin{auxlem}\label{two}
 If $f : X \to Y$ is constant and $x_1,x_2 : X$ are points, then $\mapfunc f : \id[X] {x_1}{x_2} \to \id[Y]{f(x_1)}{f(x_2)}$ is constant. In particular, $\mapfunc f$ maps every loop around $x$ (that is, path from $x$ to $x$) to $\refl {{f(x)}}$. 
\end{auxlem}
\proof
 If $c$ is the proof of $\const f$, then $\mapfunc f$ maps a path $p : \id x y$ to ${\opp{c(x,x)}} \ct {c(x,y)}$. 
 This is easily seen to be correct for $(x, x, \refl x)$, which is enough to apply path induction. As the expression is independent of $p$, the function $\mapfunc f$ is constant. The second part follows from the fact that $\mapfunc f$ maps $\refl x$ to $\refl {f(x)}$.
\qed
With these lemmata at hand, we give a proof of the Fixed Point Lemma: 
\proof[Proof of Lemma~\ref{fixedpoint}]
 Assume $f : X \to X$ is a function and $c : \const f$ is a proof that it is constant. For any two pairs $(x, p)$ and $(x', p') : \fix f$, we need to construct a path connection them. 
 
 First, we simplify the situation by showing that we can assume that $x$ and $x'$ are the same: By composing $p : \id{x}{f \, x}$ with $c(x,x') : \id{f(x)}{f(x')}$ and $\opp{(p')} : \id{f(x')}{x'}$, we get a path $p'' : \id{x}{x'}$. By a standard lemma \cite[Theorem 2.7.2]{HoTTbook}, a path between two pairs corresponds to two paths: One path between the first components, and one between the second, where transporting along the first path is needed. 
 We therefore now get that $(x, {\opp{(p'')}} \ct {p'})$ and $(x', p')$ are equal: $p''$ is a path between the first components, which makes the second component trivial. Write $q$ for the term ${\opp{(p'')}} \ct {p'}$. 
 
 We are now in the (nicer) situation that we have to construct a path between $(x, p)$ and $(x, q) : \fix f$. Again, such a path can be constructed from two paths for the two components. Let us assume that we use some path $t : \id x x$ for the first component. We then have to show that $\trans t p$ equals $q$. 
 In the situation with $(x,p)$ and $(x', p')$, it might have been tempting to use $p''$ as a path between the first components, and that would correspond to choosing $\refl x$ for $t$. However, one quickly convinces oneself that this cannot work in the general case.
 
 By Auxiliary Lemma~\ref{one}, with the identity for $h$ and $f$ for $k$, the first of the two terms, i.\,e. $\trans t p$, corresponds to ${\opp t} \ct {p} \ct {\mapfunc f t}$. 
 With Auxiliary Lemma~\ref{two}, that term can be further simplified to ${\opp t} \ct {p}$. 
 What we have to prove is now just $\id{{\opp t} \ct {p}} q$, so let us just choose ${p} \ct {\opp q}$ for $t$, thereby making it into a straightforward application of the standard lemmata.
\qed

A more elegant but possibly less revealing proof of the Fixed Point Lemma was given by Christian Sattler:
\proof[Second Proof of Lemma~\ref{fixedpoint} (Sattler)]
  Given $f : X \to X$ and $c : \const f$ as before, assume $(x_0, p_0) : \fix f$. 
 For any $x:X$, we have an equivalence of types,
 \begin{equation}
  \id{f(x)}{x} \;\, \simeq \;\, \id{f(x_0)}{x},
 \end{equation}
 given by precomposition with $c(x_0,x)$.
 Therefore, we also have the equivalence
 \begin{equation}
  \sm{x:X} \id{f(x)}{x} \;\, \simeq \;\, \sm{x:X} \id{f(x_0)}{x}.
 \end{equation}
 The second of these types is a singleton and thus contractible,
 while the first is just $\fix f$. This shows that any other inhabitant of $\fix f$ is indeed equal to $(x_0, p_0)$.
\qed
 
We will exploit Lemma~\ref{fixedpoint} in different ways. 
For the following corollary note that, given an endomap $f$ on $X$ with constancy proof $c$, we have a canonical projection 
\begin{equation}
 \fst : \fix f \to X
\end{equation}
and a function 
\begin{align} 
 &\epsilon : X \to \fix f   \label{eq:fixf-X-equiv} \\
 &\epsilon (x) \defeq \left(f(x) \, , \, c(x,f(x))\right).  \label{eq:def-epsilon}
\end{align}
\begin{cor} \label{cor:fixistrunc}
 In basic MLTT, for a type $X$ with a constant endofunction $f$, the type $\fix f$ is a proposition that is logically equivalent to $X$. 
 In particular, $\fix f$ satisfies the conditions \ref{item:trunc1}--\ref{item:trunc3} of Definition~\ref{def:htruncation}. 
 Therefore, for a type with a constant endomap, the weak propositional truncation is actually definable.
 If $\brck-$ is part of the theory, $\brck X$ and $\fix f$ are equivalent, $\eqv{\brck X}{\fix f}$. 
 \qed
\end{cor}

We are now in the position to prove the statement that we have announced at the beginning of the section.
\begin{thm}\label{thm:maintheorem}
 A type $X$ has a constant endomap if and only if it has split support in the sense that $\brck X \to X$.
\end{thm}

\proof
 As already mentioned in earlier, the ``if''-part is simple: given $\brck X \to X$, we just need to compose it with $\bproj{-} : X \to \brck X$ to get a constant endomap.
 The other direction is an immediate consequence of Corollary~\ref{cor:fixistrunc}.
\qed

We want to add the remark that $\coll X$ can be replaced by a seemingly weaker assumption.
The following statement (together with the Theorem~\ref{thm:maintheorem}) shows that it is enough to have $f : X \to X$ which is \emph{merely} constant:
\begin{thm} \label{thm:const-proof-hidden}
 For a type $X$, the following are logically equivalent:
 \begin{enumerate}
  \item $X$ has a constant endomap \label{item:trunc-const-1}
  \item $X$ has an endomap $f$ with a proof $\brck{\const f}$. \label{item:trunc-const-3}
 \end{enumerate} 
\end{thm}

\noindent The first direction is trivial, but its reversibility is interesting.
We do \emph{not} think that $\brck{\const f}$ allows us to construct an element of $\const f$.
\proof [Proof of the nontrivial direction of Theorem~\ref{thm:const-proof-hidden}]
Assume $f$ is an endofunction on $X$.
From Lemma~\ref{fixedpoint}, we know that
\begin{equation}
 \const f \to \istype {-1} (\fix f).
\end{equation}
Using the recursion principle with the fact that the statement $\istype {-1} (\fix f)$ is a proposition itself yields
\begin{equation} \label{eq:const-trunc-shows-fix-prop}
 \brck{\const f} \to \istype {-1} (\fix f).
\end{equation}

Previously, we have constructed a map 
\begin{equation}
 \const f \to \brck X \to \fix f.
\end{equation}
Let us write this function as
\begin{equation}
 \brck X \to \const f \to \fix f.
\end{equation}
This makes it trivial to define a function 
\begin{equation}
 \brck X \times \brck {\const f} \to \const f \to \fix f.
\end{equation}
We assume $\brck X \times \brck {\const f}$. From \eqref{eq:const-trunc-shows-fix-prop}, we conclude that $\fix f$ is a proposition.
Therefore, we may apply the recursion principle of the truncation and get
\begin{equation}
 \brck X \times \brck {\const f} \to \brck {\const f} \to \fix f,
\end{equation}
which, of course, gives us
\begin{equation} \label{eq:brck-fix}
 \brck X \to \fix f
\end{equation}
under the assumption~\ref{item:trunc-const-3} of the theorem. Composing $\bproj -$ with \eqref{eq:brck-fix} and with the first projection, we get a constant function $g : X \to X$.
\qed

It seems to be impossible to show that the constructed function $g$ is equal to $f$. On the other hand, it is easy to prove the truncated version of this statement:
\begin{equation}
 \brck{\prd{x:X} \id{f x}{g x}}.
\end{equation}
The detailed proof can be found in our formalization~\cite{krausEscardoEtAll_existenceFormalisation}.

\section{Factoring weakly constant Functions} \label{sec4c:factorizing}

In Theorem~\ref{thm:maintheorem} we have seen that a type $X$ with a constant function $f : X \to X$ always has split support. 
In fact, what we have done is actually slightly more: the constructed map $\overline f : \brck X \to X$ has the property that the triangle
\begin{center}
\begin{tikzpicture}[x=\textwidth/120,y=\textwidth/120]
\node (A) at (0,0) {$X$}; 
\node (B) at (30,0) {$X$}; 
\node (P) at (15,-7) {$\brck X$}; 
\draw[->] (A) to node [above] {$f$} (B);
\draw[->] (A) to node [below left] {$\bproj -$} (P);
\draw[->] (P) to node [below right] {$\overline f$} (B);
\end{tikzpicture}
\end{center}
commutes pointwise (in the sense that we have a family of equality proofs). 

It seems a natural question to ask whether the fact that $f$ is an endofunction is required: given a (weakly) constant function $f : X \to Y$, can it be factored in this sense through $\brck X$?
With Theorem~\ref{thm:maintheorem} in mind, it may be surprising that the answer is negative.
In the presence of univalence, Shulman has constructed a family of weakly constant functions such that it is impossible that all of them factor~\cite{shulman:wconst}.
From another result by the first-named author, it follows that functions $\brck X \to Y$ can be constructed from \emph{coherently} constant functions $X \to Y$, where the proof of weak constancy comes with a tower of coherence conditions~\cite{kraus_generaluniversalproperty}.
However, there are special cases in which the factorization is possible only assuming weak constancy, and some of these are discussed in the current section.

Let us start by giving a precise definition.
\begin{defi}
 Given a function $f : X \to Y$ between two types, we say that $f$ 
 \emph{factors} 
 through a type $Z$ if there are functions $f_1 : X \to Z$ and $f_2 : Z \to Y$ such that
 \begin{equation}
  \prd{x:X} \; \id[Y]{f_2(f_1(x))}{f(x)}.
 \end{equation}
 In particular, we say that $f$ factors through $\brck X$ if there is a function $\overline f : \brck X \to Y$ such that
 \begin{equation}
  \prd{x:X} \; \id[Y]{\overline{f} (\bproj{x})}{f(x)}.
 \end{equation}
\end{defi}
As we will discuss later, assuming judgmental computation for $\brck-$, a factorization in the above sense allows us to construct a judgmental factorization (see Section~\ref{sec9:judgm-beta}).

A related known result is that any function $f : X \to Y$ factors through its \emph{image} (see \cite[Chapter 7.6]{HoTTbook}), where the image $\image(f)$ is defined as
\begin{equation}
 \image(f) \defeq \sm{y : Y} \brck{\sm{x:X} \id[Y]{f(x)}{y}}. 
\end{equation}
If $\image(f)$ is propositional, this answers positively the question that we want to discuss.
We will see that this is what happens if $Y$ is a set and $f$ is constant (Theorem~\ref{thm:factor-set}).
However, in general, $\image(f)$ is not necessarily propositional even if $f$ is constant:
One can check easily that $Y$ is a set if and only if all the functions $\unit \to X$ have a propositional image (which of course means that all those images are contractible).

Constructing a function out of the propositional truncation of a type is somewhat tricky. 
A well-known \cite[Chapter 3.9]{HoTTbook} 
strategy for defining a map $\brck X \to Y$ is to construct a proposition $P$ together with functions $X \to P$ and $P \to Y$. We have already implicitly done this in previous sections. We can make this method slightly more convenient to use if we observe that $P$ does not need to be a proposition, but it only needs to be a proposition under the assumption that $X$ is inhabited:
\begin{principle} \label{lifting-principle}
 Let $X, Y$ be two types. Assume $P$ is a type such that $P \to Y$. If $X$ implies that $P$ is contractible, then there is a function $\brck X \to Y$. In particular, if $f : X \to Y$ is a function that factors through $P$, then $f$ factors through $\brck X$. 
\end{principle}
Let us briefly justify this principle. Assume that $P$ has the assumed property. Utilizing that the statement that $P$ is contractible is propositional itself, we see that $\brck X$ is sufficient to conclude that $P$ is a proposition. This allows us to prove $\brck X \times P$ to be propositional. The map $P \to Y$ clearly gives rise to a map $\brck X \times P \to Y$, and the map $X \to \brck X \times P$ is given by $\bproj-$ and the fact that $P$ is contractible under the assumption $X$. \qed

There are several situations in which this principle can be applied.
The following theorem does not need it as it is mostly a restatement of our previous result from Section~\ref{sec4:coll}.
\begin{thm} \label{thm:factorendo}
 A weakly constant function $f : X \to Y$ factors through $\brck X$ in any one of the following cases, of which the equivalent~\ref{item:3} and~\ref{item:4} generalize all others:
 \begin{enumerate}
  \item $X$ is empty, i.e.\ $X \to \emptyt$ \label{item:1}
  \item $X$ is inhabited, i.e.\ $\unit \to X$ \label{item:2}
  \item $X$ has split support, i.e.\ $\brck X \to X$ \label{item:3}
  \item $X$ has a weakly constant endofunction, i.e.\ $\sm{f:X \to X}\const f$ \label{item:4}
  \item we have any function $g : Y \to X$. \label{item:5}
 \end{enumerate}
\end{thm}
\proof
Each of \ref{item:1} and \ref{item:2} let us conclude \ref{item:3}. Further, \ref{item:5} gives us \ref{item:4} as the composition $g \circ f$ is a constant endofunction on $X$. 
The logical equivalence of~\ref{item:3} and~\ref{item:4} is Theorem~\ref{thm:maintheorem}.
Thus, it is sufficient to prove the statement for \ref{item:3}, so assume $s : \brck X \to X$. The required conclusion is then immediate 
as $f$ is pointwise equal to the composition of $\bproj - : X \to \brck X$ and $f \circ s$.
\qed

Our next statement implies what we mentioned at the beginning of Section~\ref{sec4c:factorizing}: under the assumption of unique identity proofs, the factorization is always possible.
\begin{thm} \label{thm:factor-set}
 Let $X, Y$ be again two types and $f :X \to Y$ a constant function. If $Y$ is a set, then $f$ factors through $\brck X$.
\end{thm}
\proof
The crucial observation is that the image of $f$ is propositional in the considered case.
In detail, we proceed as follows.
We define $P$ to be the image of $f$, that is,
 \begin{equation}
  P \defeq \sm{y : Y} \brck{\sm{x:X} \id[Y]{f(x)}{y}}. 
 \end{equation}
In order to apply Principle~\ref{lifting-principle}, we need to know that $f$ factors through $P$.
This is obvious from the following diagram:
\begin{center}
\begin{tikzpicture}[x=\textwidth/80,y=\textwidth/80]
\node (A) at (0,0) {$X$}; 
\node (B) at (30,0) {$Y$}; 
\node (P) at (15,-7) {$P$}; 
\draw[->] (A) to node [above] {$f$} (B);
\draw[->] (A) to node [below left] {$\lam x (f(x), \bproj{x,\refl{f(x)}})$} (P);
\draw[->] (P) to node [below right] {$\fst$} (B);
\end{tikzpicture}
\end{center}
 We need to prove that $P$ is propositional. 
 That is, given two elements $(y_1, p_1)$ and $(y_2, p_2)$ in $P$, we want to show that they are equal. 
 Let us once more construct the equality via giving a pair of paths. 
 For the second component, there is nothing to do as $p_1$ and $p_2$ live in propositional types. 
 To show $\id[Y]{y_1}{y_2}$, observe that this type is propositional as $Y$ is a set and we may thus assume that we have inhabitants $(x_1, q_1) : \sm{x_1:X} \id[Y]{f(x_1)}{y_1}$ and $(x_2, q_2) : \sm{x_2:X} \id[Y]{f(x_2)}{y_2}$ instead of $p_1$ and $p_2$. 
 But $\id{f(x_1)}{f(x_2)}$ by constancy, and therefore $\id{y_1}{y_2}$. 
 The maps $X \to P$ and $P \to Y$ are the obvious ones and the claim follows by Principle~\ref{lifting-principle} (or rather the preceding comment, the strengthened version is not needed).
\qed
It is not hard to see that, assuming function extensionality, the implication of Theorem~\ref{thm:factor-set} gives rise to an equivalence
\begin{equation}
 \eqvspace {\left( \sm{f:X\to Y} \const f \right)}{\left( \brck X \to Y\right) },
\end{equation}
where we use in particular that $\const f$ is propositional under the given conditions. 
This is the simplest non-trivial special case of the result that functions $\brck X \to Y$ correspond to \emph{coherently} constant functions $X \to Y$~\cite{kraus_generaluniversalproperty}.

Our last example of a special case in which the factorization can be done is more involved. However, it is worth the effort as it provides valuable intuition and an interesting application, as we will discuss below. The proof we give benefits hugely from a simplification by Sattler who showed to us how reasoning with type equivalences can be applied here.

\begin{thm} \label{thm:factor-coprod}
 Assume that function extensionality holds.
 If $f: X \to Y$ is constant and $X$ is the coproduct of two propositions, then $f$ factors through $\brck X$. 
\end{thm}
\proof 
Assume $X \jdeq Q + R$, where $Q$ and $R$ are propositions, and assume that $c : \const f$ is the witness of constancy. 
Define $P$ to be the following $\Sigma$-type with four components:
\begin{equation} 
 \begin{alignedat}{2} \label{eq:P-type}
  P \; \defeq \; & \Sigma && \left( y : Y \right) \\ 
           & \Sigma && \left( s :  \prd {q : Q} \; \id[]{y}{f(\inl\, q)} \right) \\
           & \Sigma && \left( t :  \prd {r : R} \; \id[]{y}{f(\inr\, r)} \right) \\
           &  && \left( \prd {q : Q} \prd {r : R} \; \idspace {\opp {s(q)} \ct {t(r)}} {c(\inl\, q \, , \, \inr\, r)} \right)
 \end{alignedat}
\end{equation}

In order to apply Principle~\ref{lifting-principle} we need to construct a function $P \to Y$ and a proof that $X$ implies that $P$ is contractible. 

The function $P \to Y$ is, of course, given by a simple projection. 
For the other part, let a point of $X$ be given. 
Without loss of generality, we assume that this inhabitant is $\inl \, q_0$ with $q_0 : Q$.
It would be possible to construct a point in $P$ and show that this point is equal to any other point. 
However, constructing a chain of equivalences yields a more elegant proof.
This strategy was proposed to us by Christian Sattler.

As $Q$ is contractible with center $q_0$, it suffices to only consider $q_0$ instead of quantifying over all elements of $Q$. 
Applying this twice shows that $P$ is equivalent to the following type:
\begin{equation}
 \begin{alignedat}{2}
  \phantom{P \; \defeq \;} & \Sigma && \left( y : Y \right) \\
           & \Sigma && \left( s :  \id[]{y}{f(\inl\, q_0)} \right) \\
           & \Sigma && \left( t :  \prd {r : R} \; \id[]{y}{f(\inr\, r)} \right) \\
           &  && \left( \prd {r : R} \; \idspace {\opp {s} \ct {t(r)}} {c(\inl\, q_0 \, , \, \inr\, r)} \right).
 \end{alignedat}
\end{equation}
The first two $\Sigma$-components together have the shape of a singleton, showing that this part is contractible with the canonical inhabitant $(f(\inl\, q_0) , \refl{})$. 
We may thus remove these $\Sigma$-components (see~\cite[Theorem 3.11.9 (ii)]{HoTTbook}) and the above type further simplifies to
\begin{equation}
 \begin{alignedat}{2}
  \phantom{P \; \defeq \;} &  \Sigma && \left( t :  \prd {r : R} \; \id[]{f(\inl\, q_0)}{f(\inr\, r)} \right) \\
           &  && \left( \prd {r : R} \; \idspace {\opp {\refl{}} \ct {t(r)}} {c(\inl\, q_0 \, , \, \inr\, r)} \right).
 \end{alignedat}
\end{equation}
We apply the distributivity principle of $\Pi$ and $\Sigma$ (see~\cite[Theorem 2.15.7]{HoTTbook}), together with standard simplifications, to further simplify to
\begin{equation}
 \begin{alignedat}{2}
  \phantom{P \; \defeq \;} &  \prd{r : R}  && \Sigma \left( t : \id[B]{f(\inl\, q_0)}{f(\inr\, r)} \right) \\
           &  && \phantom{\Sigma} \left( \idspace {{t}} {c(\inl\, q_0 \, , \, \inr\, r)} \right).
 \end{alignedat}
\end{equation}
For any $r:R$, the dependent pair part is contractible as it is, once more, a singleton, and function extensionality allows us to conclude the stated result.
\qed 
Theorem~\ref{thm:factor-coprod} was inspired by a discussion on the \emph{homotopy type theory mailing list}~\cite{hott:mailinglist}. 
Shulman observed that, for two propositions $Q$ and $R$, their \emph{join} $Q*R$ \cite[Chapter 6.8]{HoTTbook}, defined as the (homotopy) pushout of the diagram $Q \xleftarrow{\fst} Q \times R \xrightarrow{\snd} R$, is equivalent to $\brck {Q + R}$. 
This means that, in the presence of \emph{higher inductive types}~\cite[Chapter 6]{HoTTbook},
the type $\brck {Q+R}$ has the (seemingly) stronger elimination rule of the join. 
The second named author then asked whether higher inductive types do really improve the elimination properties of $\brck {Q+R}$ in this sense. 
This was discussed shortly before we could answer the question negatively with the result of Theorem~\ref{thm:factor-coprod}:
its statement about $\brck {Q+R}$ corresponds exactly to the elimination property of $Q * R$.
Thus, the join of two propositions already exists in a minimalistic setting that involves truncation but no other higher inductive types.

\section{Populatedness}\label{sec5:populatedness}

In this section we discuss a notion of \emph{anonymous existence},
similar to, but weaker (see Section~\ref{sec72:pop-inh}) than propositional truncation. It
crucially depends on the Fixed Point Lemma~\ref{fixedpoint}.
Let us start by discussing another perspective on what we have
explained in Section~\ref{sec4:coll}.

Trivially, for a type $X$, we can prove the statement
\begin{equation}\label{trivial}
 \brck X \to (\brck{X} \to X) \to X.
\end{equation}
By Lemma~\ref{thm:maintheorem}, this is equivalent to
\begin{equation}
 \brck X \to \coll X \to X,
\end{equation}
and hence
\begin{equation}\label{lesstrivial}
 \coll X \to \brck X \to X,
\end{equation}
which can be read as: If we have a constant endomap on $X$ and we wish to get an inhabitant of $X$ (or, equivalently, a fixed point of the endomap), then $\brck X$ is sufficient to do so. We can additionally ask whether it is also necessary: can we replace the first assumption $\brck{X}$ by something weaker? Looking at formula~\ref{trivial}, it would be natural to conjecture that this is not the case, but it is. 
In this section, we discuss what it can be replaced by, and in Section~\ref{sec72:pop-inh}, we give a proof that it is indeed weaker.

For answering the question what is needed to get from $\hstable X$ to $X$, let us define the following notion:
\begin{defi}[populatedness]\label{populatedness}
 For a given type $X$, we say that $X$ is populated, written~$\populated{X}$, if
every constant endomap on $X$ has a fixed point:
 \begin{equation}
  \populated{X} \defeq \fa f {X \to X} \const f \to \fix f,
 \end{equation}
where $\fix f$ is the type of fixed points, defined as in Lemma~\ref{fixedpoint}.
\end{defi}
The notion of populatedness (which, to add a caveat, is not functorial; see Theorem~\ref{tfae2})
allows us to comment on the question raised above.  
If $\populated X$ has an element and $X$ has a constant endomap, then $X$ has an
inhabitant, as such an inhabitant can be extracted from the type of
fixed points by projection. Hence, $\populated X$ instead of
$\brck X$ in~\ref{lesstrivial} would be sufficient as well.
Therefore, 
\begin{equation} \label{eq:pophstable}
 \populated X \to (\brck X \to X) \to X. 
\end{equation}
At this point, we have to ask ourselves whether~\eqref{eq:pophstable} is an improvement over~\eqref{lesstrivial}.
But indeed, we have the following property:
\begin{thm} \label{thm:trunc-to-pop}
Any merely inhabited type is populated. That is, for a type $X$, we have
\begin{equation} \label{eq:trunc-to-pop}
 \brck X \to \populated X.
\end{equation}
\end{thm}
\proof
Assume $f$ is a constant endofunction on $X$. The claim follows directly from Corollary~\ref{cor:fixistrunc}.
\qed
In Section~\ref{sec6:taboos} we will see that $\populated X$ is in fact strictly weaker than $\brck X$.

In the presence of propositional truncation, we can give an alternative characterization of populatedness. Recall that we indicate propositional truncation with the attribute \emph{merely}.
\begin{lem} \label{lem:popishstabletoh}
 In MLTT with propositional truncation, a type is populated if and only if the statement that it merely has split support implies that it is merely inhabited, or equivalently, if and only if the statement that $X$ has split support allows the construction of an element of $X$. Formally, the following types are logically equivalent:
 \begin{enumerate}
  \setlength{\itemsep}{4pt}
  \item $\populated X$ \label{item:popchar-1}
  \item $\bbrck{\brck X \to X} \to \brck X$ \label{item:popchar-2}
  \item $(\brck X \to X) \to X$. \label{item:popchar-3}
 \end{enumerate}
\end{lem}
\proof
We have already discussed \ref{item:popchar-1} $\Rightarrow$ \ref{item:popchar-3} above, see~\eqref{eq:pophstable}.
\ref{item:popchar-3} $\Rightarrow$ \ref{item:popchar-2} follows from the functoriality of the truncation operator.
For \ref{item:popchar-2} $\Rightarrow$ \ref{item:popchar-1}, assume we have a constant endofunction $f$ on $X$. Hence, we have a function $\brck X \to X$, thus $\bbrck{\brck X \to X}$ and, by assumption, $\brck X$.
But $\brck X$ is enough to construct a fixed point of $f$ by Corollary~\ref{cor:fixistrunc}.
\qed

A nice feature of the notion of populatedness is that it is definable in MLTT, and it can thus be used without making further assumptions.
For the rest of this section, let us explicitly not assume that the type theory has propositional truncations.
We can give one more characterization of populatedness, and a strong
parallel to mere inhabitance, as follows:
\begin{thm}\label{populatedLargeSmall}
 In MLTT, a type $X$ is populated if and only if any proposition that is logically equivalent to it holds,
  \begin{equation}
   \populated{X} \; \Leftrightarrow \; \prdtypes{P} \istype {-1} P \to (P \to X) \to (X \to P) \to P.
  \end{equation}
\end{thm}
Note that the only difference to the type expression in Theorem~\ref{hinhabitedLargeSmall} is that we only quantify over \emph{sub-propositions} of $X$, i.\,e. over those that satisfy $P \to X$, while we quantify over all propositions in the case of $\brck X$. 
This again shows that $\brck X$, if it exists, is at least as strong as $\populated X$.

\proof
  Let us first prove the direction ``$\rightarrow$''. Assume a proposition $P$ is given, together with functions $X \to P$ and $P \to X$. Composition of these gives us a constant endomap on $X$, exactly as in the proof of Theorem~\ref{tfae}. But then $\populated X$ makes sure that this constant endomap has a fixed point, which is (or allows us to extract) an inhabitant of $X$. Using $X \to P$ again, we get $P$. 
 
 For the direction ``$\leftarrow$'', assume we have a constant endomap $f$. We need to construct an inhabitant of $\fix f$. In the expression on the right-hand side, choose $P$ to be $\fix f$, and everything follows from Corollary~\ref{cor:fixistrunc}. 
\qed

The similarities between $\brck X$ and $\populated X$ do not stop
here. The following statement, together with the direction
``$\rightarrow$'' of the theorem that we have just proved, should be compared to the definition of $\brck X$ (that is, Definition
\ref{def:htruncation}):
\begin{thm} \label{thm:pop-like-trunc}
 For any $X$, the type $\populated X$ has the following properties:
 \begin{enumerate}
  \item $X \to \populated X$ \label{item:x-to-pop}
  \item $\istype {-1} (\populated{X})$ \qquad (if function extensionality holds).
 \end{enumerate}
\end{thm}
\proof
  The first point can be shown using the map $\epsilon$ as defined in~\eqref{eq:def-epsilon}.
  For the second, we use that $\fix f$ is a proposition (Lemma~\ref{fixedpoint}). 
  By function extensionality, a (dependent) function type is propositional if the codomain is (see~Section~\ref{sec2:preliminaries}) and we are done.
\qed

The following result, shown without using propositional truncation, is the analog to Theorem~\ref{thm:maintheorem}.
\begin{thm} \label{thm:populated-coll}
 Let $X$ be a type.
 If we have a constant endomap on $X$, 
 then $(\populated X \to X)$.
 Assuming function extensionality, this implication can be reversed.
 If $X$ is propositional, then $\coll X$ and $(\populated X \to X)$ are both inhabited (not requiring function extensionality). 
\end{thm}
\proof
 Given a constant endofunction $f$ on $X$, an inhabitant of $\populated X$ gives us $\fix f$ and thus $X$ by projection.
 For the other direction, if we have $(\populated X \to X)$, then the composition with~(Theorem~\ref{thm:pop-like-trunc}.\ref{item:x-to-pop}) gives a constant endofunction on $X$.
 If $X$ is propositional, then the identity is clearly constant.
\qed

As remarked above, $\brck -$ is an idempotent monad in an appropriate sense, while $\populated -$ is not even functorial (see Theorem~\ref{tfae2}).
However, we do have the following:
\begin{thm}
 Assuming function extensionality, the notion of populatedness is idempotent in the sense that, for a type $X$, we have an equivalence
 \begin{equation}
  \eqv{\populated {\populated X}} {\populated X}.
 \end{equation}
\end{thm}
\proof
 Theorem~\ref{thm:pop-like-trunc} shows that both sides are propositional and that there is a map ``$\leftarrow$''.
 A map ``$\rightarrow$'' is given by Theorem~\ref{thm:populated-coll}.
\qed

\section{Taboos and Counter-Models}\label{sec6:taboos}

In this section we look at the differences between the various
notions of (anonymous) inhabitance we have encountered.
We have, for a type $X$, the following chain of implications:
\begin{equation}
 X \, \Rightarrow \, \brck X \, \Rightarrow \, \populated X \, \Rightarrow \, \neg\neg X. \label{eq:chain}
\end{equation}
The first implication is trivial and the second is given by Theorem~\ref{thm:trunc-to-pop}. Maybe somewhat surprisingly, the last implication does not require function extensionality, as we do not need to prove that $\neg\neg X$ is propositional: to show 
\begin{equation}
 \populated X \to \neg\neg X \; ,
\end{equation}
let us assume $f : \neg X$. But then, $f$ can be composed with the unique function from the empty type into $X$, yielding a constant endomap on $X$, and obviously, this function cannot have a fixed point in the presence of $f$. Therefore, the assumption of $\populated X$ would lead to a contradiction, as required.

Under the assumption of $\LEM{}$, all implications of the chain~\eqref{eq:chain} except the first can be reversed as it is easy to show
\begin{equation}
 \prdtypes X (\brck X + \neg \brck{X}) \to \neg\neg X \to \brck X. 
\end{equation}
Constructively, none of the implications of~\eqref{eq:chain} should be reversible. 
To make that precise, we use what we call \emph{taboos},
showing that the provability of a statement would imply the provability of another better understood statement which is known to be not provable. 
A taboo is essentially a type-theoretic \emph{Brouwerian counterexample} (``constructive taboo'') or a homotopical analog (``homotopical taboo'').

In this section, we present the following discussions:
\begin{enumerate}
 \item 
 We start by assuming that the first implication can be reversed, i.e.\ that we have a function $\prdtypes X \brck X \to X$.
 It is easy to see that this assumption implies that all types are sets.
 We show the more interesting result that all equalities are decidable. 
 As an additional argument, if every type has split support, a form of choice that does not belong to type theory is implied.
 Moreover, we observe that $\brck X \to X$ can be read as ``the map $\bproj - : X \to \brck X$ is a split epimorphism'' (where the latter notion must be read with care), and we show that already the weaker assumption that it is an epimorphism implies that all types are sets.
 \item It would be nice if the second implication could be reversed, as this would imply that propositional truncation is definable in MLTT. 
 However, this is logically equivalent to a certain weak version of the axiom of choice discussed below, which is not provable (but holds under $\LEM {}$).
 \item Assuming function extensionality,
 the last implication can be reversed if and only if $\LEM {}$ holds.
\end{enumerate}

\subsection{Inhabited and Merely Inhabited}  \label{subsec:pure-trunc}

We first examine the question whether the first part of the chain~\eqref{eq:chain} can be reversed.
If $X$ is a type, it is weaker to have an inhabitant of $\brck X$ than to have an inhabitant of $X$.
It is unreasonable to expect that we can show in type theory that every type has split support, but it is interesting to see what it would imply.

First of all, if we assume that all types have split support, then this in particular holds for path spaces, and by Theorem~\ref{tfae}, every type is a set.
This assumption also implies the axiom of choice~\cite[Chapter 3.8]{HoTTbook}. 
If we have univalence for propositions and set quotients, this allows us to use Diaconescu's proof of $\LEM {}$ (\cite{Diaconescu}, see~\cite[Theorem 10.1.14]{HoTTbook}).
We want to present a similar construction in the much more minimalistic theory that we consider in the current article.

Using Theorem~\ref{thm:maintheorem}, we can formulate the assumption that all types have split support without using truncations as ``every type has a constant endofunction'',
\begin{equation}\label{eq:global-coll}
 \prdtypes X \coll X.
\end{equation}
From a constructive point of view, this is an interesting assumption. It clearly follows from $\LEM \infty$:
if we know an inhabitant of a type, we can immediately construct a constant endomap, and for the empty type, considering the identity function is sufficient.
The assumption~\eqref{eq:global-coll} contains some form of choice, but we do not expect that the general principle $\LEM\infty$ can be derived in our setting.
Hence, we may understand~\eqref{eq:global-coll} as a weak form of $\LEM\infty$. 
However, what we can derive is $\LEM\infty$ for all path spaces, i.e.\ that all types are discrete, see Lemma~\ref{martin:all-collapsible-discrete} and Theorem~\ref{theorem:coll-discrete} below.

\begin{lem}\label{martin:all-collapsible-discrete}
 In basic MLTT (without function extensionality and without propositional truncations), let $A$ be a type and $a_0, a_1 : A$ two points. 
 If the type ${(\id{a_0}{x}) + (\id{a_1}{x})}$ has a constant endomap for all $x : A$ , then $\id{a_0}{a_1}$ is decidable.
\end{lem}
As we will see in the proof, we need to know $\bfalse \not=_{\bool} \btrue$ for Lemma~\ref{martin:all-collapsible-discrete}, which can be proved using a universe.
If we assume $\bfalse \not=_{\bool} \btrue$, the lemma is true in an even weaker setting without a type universe.
Before giving the proof of Lemma~\ref{martin:all-collapsible-discrete}, we state an immediate corollary: 
\begin{thm}\label{theorem:coll-discrete}
 If every type has a constant endofunction then every type has decidable equality,
 \begin{equation}
  (\prdtypes X \coll X) \to \prdtypes X \deceq X.
 \end{equation} \qed
\end{thm}
\proof[Proof of Lemma~\ref{martin:all-collapsible-discrete}]
For (technical and conceptual) convenience, we regard the elements $a_0, a_1$ as a single map 
\begin{equation}
 a : \bool \to A 
\end{equation}
and we use 
\begin{equation}
 E_x \defeq \sm{i : \bool} \; \id{a_i}{x}  
\end{equation}
in place of the type ${(\id{a_0}{x}) + (\id{a_1}{x})}$. 
In a theory with propositional truncation, the \emph{image} of $a$ can be defined to be $\sm{x:A} \brck {E_x}$ \cite[Definition 7.6.3]{HoTTbook}.
By assumption, we have a family of constant endofunctions $f_x$ on $E_x$, and by the discussion above, we can essentially regard the type
\begin{equation}
 E \defeq \sm{x:A} \fix f_x,
\end{equation}
which can be unfolded to
\begin{equation}
 \sm{x:A} \sm{(i,p) : E_x} \id{f_x(i,p)}{(i,p)},
\end{equation}
as the image of $a$. 
It is essentially the observation that we can define this image that allows us to mimic Diaconescu's argument.
Recall from~\eqref{eq:def-epsilon} that $\epsilon$ is the canonical function that maps a point of a type to a fixed point of a given endofunction on that type. 
Clearly, $a$ induces a map
\begin{align}
 &r : \bool \to E \\
 &r(i) \defeq (a_i , \epsilon(i, \refl{a_i})).
\end{align}
Using that the second component is an inhabitant of a proposition, we have 
\begin{equation} \label{eq:ar}
 \id{r(i)}{r(j)} \, \Leftrightarrow \, \id{a_i}{a_j}.   
\end{equation}
The type $E$ can be understood as the quotient of $\bool$ by the equivalence relation $\sim$, given by $i \sim j \jdeq \id{a_i}{a_j}$.
If $E$ was the image of $a$ in the ordinary sense~\cite[Definition 7.6.3]{HoTTbook}, the axiom of choice would be necessary to find a section of $r$ (see~\cite[Theorem 10.1.14]{HoTTbook}). In our situation, this section is given by a simple projection,
\begin{align}
 &s : E \to \bool \\
 &s(x, ((i,p) , q)) \defeq i.
\end{align}
It is easy to see that $s$ is indeed a section of $r$ in the sense of $\fa{e}{E}\id{r(s(e))}{e}$. Given $(x, ((i,p),q)) : E$, applying first $s$, then $r$ leads to $(a_i, \epsilon(i, \refl{a_i}))$. 
Equality of these expressions is equality of the first components due to the propositional second component. But $p$ is a proof of $\id{a_i}{x}$. 
From that property, we can conclude that, for any $e_0, e_1:E$,
\begin{equation}\label{eq:ere}
 \id{e_0}{e_1} \, \Leftrightarrow \, \id{s(e_0)}{s(e_1)}.
\end{equation}
Combining \eqref{eq:ar} and \eqref{eq:ere} yields
\begin{equation}
 \id{a_i}{a_j} \, \Leftrightarrow \, \id{s(r(i))}{s(r(j))},
\end{equation}
where the right-hand side is an equality in $\bool$ and thus always decidable. In particular, $\id{a_0}{a_1}$ is hence decidable.
\qed

Another consequence of the assumption~\eqref{eq:global-coll} is a form of 
choice that does not belong to intuitionistic type theory. In order
to formulate and prove this, we need a few definitions.

We say that a relation $R : X
\times X \to \UU$ is \emph{propositionally valued} if 
  \begin{equation} 
   \fapairs x y X \istype {-1} (R(x,y)).
  \end{equation}
The $R$-\emph{image} of a point $x : X$ is
\begin{equation}
R_x \defeq \sm {y:X} {R(x,y)}.
\end{equation}
We say that $R$ is \emph{functional} if its point-images are all propositions:
  \begin{equation} 
   \fa x X \istype {-1} R_x.
  \end{equation}
We say that two relations $R,S : X
\times X \to \UU$ \emph{have the same domain} if
  \begin{equation}
   \fa x X R_x \Leftrightarrow S_x,
  \end{equation}
and that $S$ is a \emph{subrelation} of $R$ if
  \begin{equation}
   \fapairs x y X S(x,y) \to R(x,y).
  \end{equation}
\begin{thm}
  If all types have constant endofunctions, then every binary relation 
  has a functional, propositionally valued
  subrelation with the same domain.
\end{thm}
\proof Assume that $R : X \times X \to U$ is given. For $x:X$, let
$k_x : R_x \to R_x$ be the constant map given by the
assumption~\eqref{eq:global-coll} that all types have constant endofunctions. 
Define further
\begin{equation}
 S (x,y) \defeq \sm {a:R(x,y)} {\id{(y, a)}{k_x(y , a)}}. 
\end{equation}
Then $S$ is a subrelation of $R$ by construction.
We observe that $S_x$ is equivalent to $\fix(k_x)$ and therefore propositional (by Lemma~\ref{fixedpoint}), proving that $S$ is functional.
Together with Corollary~\ref{cor:fixistrunc}, this further shows
\begin{equation}
 R_x \Leftrightarrow \fix{k_x} \Leftrightarrow S_x,
\end{equation}
showing that $R$ and $S$ have the same domain.

What remains to show is that $S(x,y)$ is always a proposition. 
Let $s, s' : S(x,y)$. 
As $S_x$ is propositional we know $\id[S_x]{(y,s)}{(y,s')}$. By the standard lemma this type corresponds to a dependent pair type with components
\begin{align}
 &p : \id[X]{y}{y} \\
 &q : \id[S(x,y)]{\trans p s} {s'}.
\end{align}
In our case, as every type is a set, we have $\id p {\refl y}$, and $q$ gives us the required proof of $\id[S(x,y)] {s}{s'}$.
\qed

Instead of the logically equivalent formulation~\eqref{eq:global-coll}, let us now assume the original assumption that $\bproj -$ can be reversed, that is,
\begin{equation} \label{eq:all-h-stable-2}
 \prdtypes X \brck X \to X.
\end{equation} 
Note that a map $h : \brck X \to X$ is automatically a section of $\bproj- : X \to \brck X$ in the sense of
\begin{equation}
 \fa z {\brck X} \id{\bproj {h(z)}}{z}
\end{equation}
as any two inhabitants of $\brck X$ are equal. Therefore, we may read~\eqref{eq:all-h-stable-2} as:
\begin{equation}
 \text{For any type $X$, the map $\bproj- : X \to \brck X$ is a \emph{split epimorphism}.}
\end{equation}
We want to consider a weaker assumption, namely
\begin{equation}
 \text{For any type $X$, the map $\bproj- : X \to \brck X$ is an \emph{epimorphism},}
\end{equation}
where we call $e : U \to V$ an \emph{epimorphism} if, for any type $W$ and any two functions $f,g : V \to W$, we have 
\begin{equation} \label{eq:e-is-epi-def}
 (\fasimple u U \id{f(e \, u)}{g(e \, u)}) \to \fasimple v V \id{f \, v}{g \, v}.
\end{equation}
Of course, under function extensionality, $e$ is an epimorphism if and only if, for all $W,f,g$, we have
\begin{equation}
 \id{f \circ e}{g \circ e} \to \id f g.
\end{equation}
A caveat is required. Our definition of \emph{epimorphism} is the direct naive translation of the usual $1$-categorical notion into type theory. 
However, the category of types and functions with the type of equalities is not only an ordinary category, but rather an $(\infty,1)$-category.
The definition~\eqref{eq:e-is-epi-def} makes sense in the category of sets~\cite[Chapter 10.1]{HoTTbook}, where equalities are propositional.
However, the property of being an epimorphism in our sense is not propositional and it could rightfully be argued that it might not be the ``correct'' definition in a context where not every type is a set, similarly to how we argued that $\LEM \infty$ is a problematic version of the principle of excluded middle.
Despite this, we use the notion as we think that it helps providing an intuitive meaning to the plain type expression~\eqref{eq:e-is-epi-def}.

\begin{lem} \label{lem:global-epi-to-set}
 Let $Y$ be a type. 
 If the map $\bproj- : (\id{y_1}{y_2}) \to \brck{\id{y_1}{y_2}}$ is an epimorphism for any points $y_1, y_2 : Y$, 
 then $Y$ is a set.
\end{lem}
\proof
Assume $Y, y_1, y_2$ are given. 
Define two functions
\begin{equation}
 f, g : \brck{\id{y_1}{y_2}} \to Y
\end{equation}
by
\begin{align}
 &f(q) \defeq y_1,\\
 &g(q) \defeq y_2,
\end{align}
that is, $f$ and $g$ are constant at $y_1$ and $y_2$, respectively.

With these concrete choices, our assumption~\eqref{eq:e-is-epi-def} with $e \equiv \bproj-$ becomes
\begin{equation}
 \left({\id{y_1}{y_2}} \to \id{y_1}{y_2}\right)  \to  \left(\brck{\id{y_1}{y_2}} \to \id{y_1}{y_2}\right)
\end{equation}
which, of course, gives us a function
\begin{equation}
 \brck{\id{y_1}{y_2}} \to \id{y_1}{y_2}.
\end{equation}
The statement of the lemma then follows from Theorem~\ref{tfae}.
\qed

The following result summarizes the statements of Theorem~\ref{theorem:coll-discrete} and Lemma~\ref{lem:global-epi-to-set}: 
\begin{thm}
 In basic MLTT with weak propositional truncation,
 \begin{enumerate}
  \item if $\bproj - : X \to \brck X$ is a split epimorphism for every $X$, then all types have decidable equality
  \item if $\bproj - : X \to \brck X$ is an epimorphism for every $X$, then all types are sets.
 \end{enumerate}
\end{thm}
\proof
The first part is a reformulation of Theorem~\ref{theorem:coll-discrete}, while the second part is a corollary of Lemma~\ref{lem:global-epi-to-set}. \qed

\subsection{Merely Inhabited and Populated}\label{sec72:pop-inh}

Assume that the second step in~\eqref{eq:chain} can be reversed, meaning that we have
\begin{equation}
 \prdtypes X \populated X \to \brck X.
\end{equation}
Repeated use of the Fixed Point Lemma leads to a couple of interesting logically equivalent statements. 

In the previous subsection, we have discussed that we cannot show that every type has split support. However, a weaker version of this is provable:
\begin{lem}\label{pophstable}
 For every type $X$, the statement that it has split support is populated,
 \begin{equation}
  \populated{\brck X \to X}.
 \end{equation}
\end{lem}
To demonstrate the different possibilities that the logically equivalent formulations of populatedness offer, we want to give more than one proof.
The first one uses Definition~\ref{populatedness}:
\proof[First proof]
 Assume we are given a constant endofunction $f$ on $\brck X \to X$. We need to construct a fixed point of $f$, or correspondingly, any inhabitant of $\brck X \to X$. 
 By Theorem~\ref{thm:maintheorem}, a constant function $g : X \to X$ is enough for this. 
 Given $x:X$, we may apply $f$ to the function that is everywhere $x$, yielding an inhabitant of $\brck X \to X$. Applying it to $\bproj x$ gives an element of $X$, and we define $g(x)$ to be this element. The proof that that $f$ is constant immediately translates to a proof that $g$ is constant.
\qed
Alternatively, we can use the logically equivalent formulation of populatedness, proved in Theorem~\ref{populatedLargeSmall}: 
\proof[Second proof]
 Assume $P$ is a proposition and we have a proof of 
 \begin{equation}
  P \; \Leftrightarrow \; (\brck X \to X).
 \end{equation}
We need to show $P$. The logical equivalence above immediately provides an inhabitant of $X \to P$, and, by the rules of the propositional truncation, therefore $\brck X \to P$. Assume $\brck X$. We get $P$, thus $\brck X \to X$ with the above equivalence, and therefore $X$ (using the assumed $\brck X$ again). This shows $\brck X \to X$, and consequently, $P$.
\qed
Finally, we can also use that $\populated -$ can be written in terms of $\brck -$:
\proof[Third proof]
Using Lemma~\ref{lem:popishstabletoh}\ref{item:popchar-3}, the statement that needs to be shown becomes
\begin{equation}
 \left(\bbrck{\brck X \to X} \to \brck X \to X\right) \to \left(\brck X \to X\right),
\end{equation}
which is immediate.
\qed

The assumption that populatedness and mere inhabitance are equivalent has a couple of ``suspicious'' consequences, as we want to show now.
\begin{thm} \label{tfae2}
 In MLTT with weak propositional truncation, the following are logically equivalent:
 \begin{enumerate}
  \item every populated type is merely inhabited, \label{tfae:1}
  \begin{equation}
   \prdtypes X \populated X \to \brck X
  \end{equation}
  \item every type merely has split support, \label{tfae:2} 
  \begin{equation}
   \prdtypes X \bbrck {\brck X \to X}
  \end{equation}
  \item every proposition is \emph{projective} in the following sense: \label{tfae:3}
  \begin{equation}
   \prdtypes P \istype {-1} P \to \fa{Y}{P \to \UU} (\prd{p : P} \brck{Y(p)}) \to \brck{\prd{P} Y}
  \end{equation}
  (note that this is the \emph{axiom of choice} \cite[Chapter 3.8]{HoTTbook} 
  for propositions, without the requirement that $Y$ is a family of sets)
  \item $\populated{-} : \UU \to \UU$ is functorial in the sense that \label{tfae:4}
  \begin{equation}
   \fapairs{X}{Y}{\UU} (X \to Y) \to (\populated X \to \populated Y),
  \end{equation}
  where this naming is justified at least in the presence of function extensionality which implies that $\populated X \to \populated Y$ is propositional, ensuring $\id{\populated{g \circ f}}{\populated g \circ \populated f}$. 
 \end{enumerate}
 Further,~\ref{tfae:4} can be formulated in MLTT without assumptions on the availability of propositional truncation. 
 If it holds, then $\populated -$ satisfies the recursion principle of the weak propositional truncation. 
 Additionally assuming function extensionality, $\populated -$ can then serve as an implementation of the weak propositional truncation. 
\end{thm}
\proof
Let us first show the final claim. If $Y$ is propositional, then $\populated Y \to Y$ by Theorem~\ref{thm:populated-coll}. Together with \ref{tfae:4}, this gives the claimed recursion principle. The rest of the properties of the weak propositional truncation is given by Theorem~\ref{thm:pop-like-trunc}. 

Let us show the logical equivalence of the four types. The above observation immediately implies \ref{tfae:4} $\Rightarrow$ \ref{tfae:1}.
The direction \ref{tfae:1} $\Rightarrow$ \ref{tfae:4} is also immediate by functoriality of $\brck -$.
The logical equivalence of the first two points follows easily from what we already know.
\ref{tfae:1} $\Rightarrow$ \ref{tfae:2} is an application of Lemma~\ref{pophstable}, while \ref{tfae:2} $\Rightarrow$ \ref{tfae:1} follows from Lemma~\ref{lem:popishstabletoh}.

Let us now show \ref{tfae:1} $\Rightarrow$ \ref{tfae:3}. 
Let $P$ be some proposition and $Y : P \to \UU$ some family of types.
If we assume \ref{tfae:1}, it is then enough to prove
  \begin{equation}
   \left(\prd{p : P} \brck{Y(p)}\right) \to \populated{\prd{P} Y}.
  \end{equation}
By Lemma~\ref{lem:popishstabletoh}, it is enough to show
  \begin{equation} \label{eq:auxtfa}
   \left(\prd{p : P} \brck{Y(p)}\right) \to (\brck{\prd{P} Y} \to \prd{P} Y) \to \prd{P} Y.
  \end{equation}
Under several assumptions, one of them being that some $p_0 : P$ is given, we need to construct an inhabitant of $Y(p_0)$.
Recall the principle of the \emph{neutral contractible exponent} that we used in the proof of Theorem~\ref{thm:factor-coprod}.
Here, it allows us to replace $\prd{P} Y$ by $Y(p_0)$ and $\prd{p:P} \brck{Y(p)}$ by $\brck{Y(p_0)}$, and the type~\eqref{eq:auxtfa} becomes
  \begin{equation} 
   \brck{Y(p_0)} \to (\brck{Y(p_0)} \to Y(p_0)) \to Y(p_0),
  \end{equation}
which is obvious.

\ref{tfae:3} $\Rightarrow$ \ref{tfae:2} can be seen easily by
taking $P$ to be $\brck X$ and $Y$ to be constantly $X$.
\qed

Consider the third of the four statements in Theorem~\ref{tfae2}.
When $Y(p)$ is a set with exactly two elements for every $p : P$, this 
amounts to
\emph{the world's simplest axiom of choice}~\cite{wsac}, which fails in
some toposes. 
We expect that this makes it possible to show that, in MLTT with weak propositional truncation, ${\prdtypes X \populated X \to \brck X}$ is not derivable.

\subsection{Populated and Non-Empty}

If we can reverse the last implication of the chain, we have
\begin{equation}
 \prdtypes X \neg\neg X \to \populated X.
\end{equation}
To show that this cannot be provable, 
we show that it is equivalent to $\LEM {}$, a constructive taboo.
\begin{thm}
 With function extensionality, we have the following (logical) equivalence: 
\begin{equation}
  \left( \prdtypes X \neg\neg X \to \populated X \right) \; \Leftrightarrow \; \LEM{}.
\end{equation}
\end{thm}
\proof
 The direction ``$\leftarrow$'' is easy: from $X \to \populated X$, we get $\neg\neg X \to \neg\neg \populated X$.
 As $\populated X$ is propositional, $\LEM{}$ gives us $\neg\neg \populated X \to \populated X$.

 For the direction ``$\rightarrow$'', assume that $P$ is a proposition. 
 Thus, the type $P + \neg P$ is a proposition as well, and hence, the identity function on $P + \neg P$ is constant.
 
 It is straightforward to construct a proof of $\neg\neg \left(P + \neg P\right)$. By the assumption, this means that $P + \neg P$ is populated, i.e.\ every constant endomap on it has a fixed point. Therefore, we can construct a fixed point of the identity function, which is equivalent to proving $P + \neg P$.
\qed

\section{Propositional Truncation with Judgmental Computation Rule}\label{sec9:judgm-beta}

Propositional truncation is often defined to satisfy the judgmental computation rule \cite[Chapter 3.7]{HoTTbook},
 \begin{equation} 
  \elim (P, h, f, \bproj x)\; \jdeq_\beta \; f(x) \label{eq:jdgm-beta}
 \end{equation}
for any function $f : X \to P$ where $x:X$ and $P$ is propositional. 
In our discussion, we did not assume it to hold so far. 
We certainly do not want to argue that a theory without this judgmental equation is to be preferred, we simply did not need it. 
We agree with the very common view (see the introduction of \cite[Chapter 6]{HoTTbook}) that judgmental computation rules are often advantageous, not only for truncations, but for \emph{higher inductive types} \cite[Chapter 6]{HoTTbook} in general. Without them, some expressions will need to involve a ridiculous amount of transporting, just to make them type-check, and the ``computation'' will have to be done manually in order to simplify terms. If~\eqref{eq:jdgm-beta} is assumed, it suggests itself to also assume a judgmental computation rule for the induction principle, that is
 \begin{equation}  
  \depelim (P, h, f, \bproj x)\; \jdeq_\beta \; f(x), \label{eq:jdgm-beta-dep}
 \end{equation}
where $P : \brck X \to  \UU$ might now be a type family and $f: \prd{z:\brck X}P(z)$ is a dependent function rather than a simple function.
Interestingly,
it does not seem to be possible to construct $\depelim$ from $\elim$ such that~\eqref{eq:jdgm-beta-dep} holds if~\eqref{eq:jdgm-beta} holds.
In particular, the term constructed in Lemma~\ref{lem:ind-from-rec} does not have the expected judgmental computation rule.

Having said this, the judgmental $\beta$-rules do have some other noteworthy consequences. Unlike the previous results, the statements in this part of our article do need the computation rules to hold judgmentally.
So far, all our lemmata and theorems have been internal to type theory. This is only partially the case for the results from this section, as any statement that some equality holds judgmentally is a meta-theoretic property. We thus can not implement such a statement as a type in a proof assistant such as Agda, but we can still use Agda to check our claims; for example, if
\begin{align}
 &p : \id x y \\
 &p \defeq \refl x
\end{align}
type-checks, we may conclude that the equality does hold judgmentally.

\subsection{The Interval} 

The interval $\I$ as a higher inductive type \cite[Chapter 6.3]{HoTTbook} is a type in homotopy type theory that consists of two points $i_0, i_1 : \I$ and a path $\seg : \id[\I]{i_0}{i_1}$ between them.
Its \emph{recursion}, or \emph{non-dependent elimination} principle says: Given
\begin{align}
 &Y : \UU \label{eq:i-ass-1} \\
 &y_0 : Y \\
 &y_1 : Y \\
 &p : \id{y_0}{y_1}, \label{eq:i-ass-4}
\end{align}
there exists a function $f : \I \to Y$ such that 
\begin{align}
 &f(i_0) \jdeq y_0   \label{eq:i-point-1} \\
 &f(i_1) \jdeq y_1  \label{eq:i-point-2}  \\
 &\id{\mapfunc f (\seg)}{p}. \label{eq:i-seg}
\end{align}
For the interval's induction principle, we refer to~\cite[Chapter 6.3]{HoTTbook}.
The interval is a contractible type and as such equivalent to the unit type. However, this does not make it entirely boring; it is the \emph{judgmental} equalities that matter. Note that the \emph{computation rules} for the \emph{points} are judgmental (\ref{eq:i-point-1},\ref{eq:i-point-2}), while the rule for the path~\eqref{eq:i-seg} is only given by an equality proof.

We will now show that $\brck \bool$ can be regarded as the interval.
\begin{thm} \label{thm:bool-is-interval}
 For the type $\brck \bool$, the recursion principle of the interval (including the computational behavior) is derivable using~\eqref{eq:jdgm-beta}, and the induction principle follows from~\eqref{eq:jdgm-beta-dep}.
\end{thm}
\proof
 We only show that the recursion principle is derivable, which will be sufficient for the subsequent developments.
 The induction principle can be derived very similarly.
We need to show that, under the assumptions~(\ref{eq:i-ass-1}-\ref{eq:i-ass-4}), there is a function $f : \brck \bool \to Y$ such that 
\begin{align}
 &f(\bproj \bfalse) \jdeq y_0   \label{eq:bool-point-1} \\
 &f(\bproj \btrue) \jdeq y_1  \label{eq:bool-point-2}  \\
 &\id{\mapfunc f (\h_{\bproj \bfalse, \bproj \btrue})}{p}. \label{eq:bool-seg}
\end{align}
We define
\begin{align}
 &g : \bool \to \sm{y : Y} \id {y_0} y \\
 &g (\bfalse) \defeq (y_0, \refl {}) \\
 &g (\btrue) \defeq (y_1, p).
\end{align}
As $\sm{y : Y} \id {y_0} y$ is contractible, $g$ can be extended to a function $\overline g : \brck \bool \to \sm{y : Y} \id {y_0} y$, and we define $f \defeq \fst \circ \overline g$. 
It is easy to check that $f$ has indeed the required judgmental properties~\eqref{eq:bool-point-1} and~\eqref{eq:bool-point-2}.
The equality~\eqref{eq:bool-seg} is only slightly more difficult: First, using the definition of $f$ and a standard functoriality property of $\mapfunc {}$ \cite[Lemma 2.2.2 (iii)]{HoTTbook}, we observe that $\mapfunc f (\h_{\bproj \bfalse, \bproj \btrue})$ may be written as
\begin{equation}
 \mapfunc \fst (\mapfunc {\overline g} (\h_{\bproj \bfalse, \bproj \btrue})).
\end{equation}
But here, the path $\mapfunc {\overline g} (\h_{\bproj \bfalse, \bproj \btrue})$ is an equality in the contractible type $\id{(y_0, \refl {})}{(y_1 , p)}$ (note that both terms inhabit a contractible type themselves) and thereby unique. In particular, it is equal to the path which is built out of two components, the first of which can be chosen to be $p$ (the second component can then be taken to be a canonically constructed inhabitant of $\id{\trans p {\refl{}}}{p}$).
\qed

\subsection{Function Extensionality} \label{subsec:beta-funext}

It is known that the interval $\I$ with its judgmental computation rules implies function extensionality. We may therefore conclude that propositional truncation is sufficient as well.

\begin{lem}[Shulman~\cite{blog_interval}] \label{lem:int-funext}
 In a type theory with $\I$ and the judgmental $\eta$-law for functions (which we assume), function extensionality is derivable.
\end{lem}
\proof
Assume $X,Y$ are types and $f,g : X \to Y$ are functions, and $h : \fasimple x X \id{f(x)}{g(x)}$ a proof that they are pointwise equal. Using the recursion principle of $\I$, we may then define a family
\begin{equation}
 k : X \to \I \to Y \\
\end{equation}
 of functions, indexed over $X$, such that $k (x , i_0) \jdeq f(x)$ and $k (x , i_0) \jdeq g(x)$ for all $x:X$; of course, we use $h(x)$ as the required family of paths. Switching the arguments gives a function 
\begin{equation}
 k' : \I \to X \to Y \\
\end{equation}
 with the property that $k' (i_0) \jdeq f$ and $k' (i_1) \jdeq g$ (by $\eta$ for functions), and thereby $\mapfunc {k'} (\seg) : \id f g$. 
\qed

The combination of Theorem~\ref{thm:bool-is-interval} and Lemma~\ref{lem:int-funext} implies:
\begin{cor}
 In type theory with propositional truncation that satisfies the judgmental computation rule, function extensionality can be derived. 
 \qed
\end{cor}

\subsection{Judgmental Factorization}
The judgmental computation rule of $\brck -$ also allows us to factor any function \emph{judgmentally} through the propositional truncation as soon as it can be factored in any way. This observation is inspired by and a generalization of the fact that $\brck \bool$ satisfies the judgmental properties of the interval (Theorem~\ref{thm:bool-is-interval}).
\begin{thm} \label{thm:jdg-factor-nondep}
 Any (non-dependent) function that factors through the propositional truncation can be factored judgmentally: assume types $X,Y$ and a function $f : X \to Y$ between them. Assume that there is $\overline f : \brck X \to Y$ such that
 \begin{equation}
  h : \fa x X \id{f(x)}{\overline f(\bproj x)}.
 \end{equation}
Then, we can construct a function $f' : \brck X \to Y$ such that, for all $x:X$, we have
\begin{equation}
 f(x) \jdeq f'(\bproj x),
\end{equation}
which means that the type $\fa x X \id{f(x)}{f'(\bproj x)}$ is inhabited by the function that is constantly $\refl {}$.
\end{thm}
\proof
We define a function 
\begin{align}
 &g : X \to \prd{z:\brck X} \sm{y:Y} \id y {\overline f(z)} \label{eq:jdgm-fact-g-type} \\
 &g(x) \defeq \lam z \left( f(x), h(x) \ct \mapfunc {\overline f} (\h_{\bproj x, z}) \right)
\end{align}
By function extensionality and the fact that singletons are contractible, the codomain of $g$ is contractible, 
and thus, we can extend $g$ and get
\begin{equation}
 \overline g : \brck X \to \prd{z:\brck X} \sm{y:Y} \id y {\overline f(z)}. \label{eq:jdgm-fact-lift-g-type}
\end{equation}
We define 
\begin{equation} \label{eq:f'-for-judgmental}
 f' \defeq \lam {z:\brck X} \fst(\overline g \, z \, z)
\end{equation}
and it is immediate to check that $f'$ has the required properties.
\qed

Note that in the above argument we have only used~\eqref{eq:jdgm-beta}. We have avoided~\eqref{eq:jdgm-beta-dep} by introducing the variable $z$ in~\eqref{eq:jdgm-fact-g-type}, which is essentially a duplication of the first argument of the function, as it becomes apparent in~\eqref{eq:f'-for-judgmental}.

Furthermore, we have assumed that $f$ is a non-dependent function. 
The question does not make sense if $f$ is dependent in the sense of $f : \prd{x:X}Y(x)$; however, it does for $f: \prd{x : X}Y(\bproj x)$. In this case, it seems to be unavoidable to use~\eqref{eq:jdgm-beta-dep}, but the above proof still works with minimal adjustments.
We state it for the sake of completeness.

\begin{thm}
 Let $X$ be a type and $Y : \brck X \to \UU$ a type family. Assume we have functions 
 \begin{align}
  & f : \prd{x:X}{Y(\bproj x)} \\
  & \overline f : \prd{z : \brck X} Y(z)
 \end{align}
 such that 
 \begin{equation}
  \fa x X \id[Y(\bproj x)]{f(x)}{\overline f(\bproj x)}.
 \end{equation}
 Then, we can construct a function $f' : \prd{z : \brck X}B(z)$ with the property that for any $x:X$, we have the judgmental equality
 \begin{equation}
  f(x) \jdeq f'(\bproj x).
 \end{equation}
\end{thm}
\proof
Because we allow ourselves to use~\eqref{eq:jdgm-beta-dep} the proof becomes actually simpler than the proof above.
This time, we can define 
\begin{align}
 &g : \prd{x : X}  \sm{y:Y} \id y {\overline f(\bproj x)} \\
 &g(x) \defeq \left( f(x), h(x) \right).
\end{align}
Using~\eqref{eq:jdgm-beta-dep}, we get
\begin{equation}
 \overline g : \prd{z : \brck X}  \sm{y:Y} \id y {\overline f(z)}. 
\end{equation}
Then, 
\begin{equation} 
 \fst \circ \overline g 
\end{equation}
fulfils the required condition.
\qed

\subsection{An Invertibility Puzzle}

For a type $X$, the function $\bproj - : X \to \brck X$ turns an element $x : X$ into an anonymous inhabitant $\bproj x : \brck X$.
It is thus reasonable to think of $\bproj -$ as a function that hides information.
However, as we will demonstrate, this interpretation is only justified as long as we think of internal properties.
We will show that the function $\bproj -$ does not erase any meta-theoretical information in the following sense:
Assume $z : \brck X$ is defined to be $\bproj x$ for some $x : X$. 
Without looking at this definition, we can recover $x$ (e.g.\ in a proof assistant, $z$ could be imported from another file; then, we do not need to open that file in order to find out $x$).
To do this, we only need to observe how $z$ computes in a suitable environment.
To be precise, we construct a term $\myst_X$ such that, for any $z$ as above, the expression $\myst_X (z)$ is judgmentally equal to $x$.

The meta-theoretic statement that we can recover $x$ from $z$ is true, but in general, $\myst_X$ might not be a closed term (i.e.\ could depend on some assumptions which do not influence the computation).
However, assuming the univalence axiom, $\myst_X$ can be constructed without any further assumptions for a non-trivial class of types including the natural numbers.
That is, in MLTT with propositional truncations and the univalence axiom, we can construct a term $\myst_{\N}$ such that
\begin{align}
 &\mathsf{id'} : \N \to \N \\
 &\mathsf{id'} \defeq \lam n \myst_\N (\bproj{n})
\end{align}
type-checks and $\mathsf{id'}$ is the identity function on $\N$, with a proof
\begin{align}
 &\mathsf{p} : \prd{n : \N} \id{\mathsf{id'}(n)}{n} \\
 &\mathsf{p} \defeq \lam n \refl n.
\end{align}
We think that the possibility to do this is counter-intuitive and surprising. 
In particular it may seem that we could apply $\mapfunc {\myst_\N}$ on the canonical inhabitant of $\id[\brck \N]{\bproj 0}{\bproj 1}$ to conclude $\id[\N] 0 1$.
However, this would only work if the type of $\myst_\N$ was $\brck \N \to \N$, which it is not; it is a $\Pi$-type that is not easier to write down than the full definition of its inhabitant $\myst_\N$ itself.
In the following, we show the full construction. 
For further discussion, see the homotopy type theory blog entry by the first named author~\cite{kraus:pseudoinverse}, where this result was presented originally.

First, let us state two useful general definitions:
\begin{defi}[Pointed Types {{\cite[Definition~2.1.7]{HoTTbook}}}]
A \emph{pointed type} is a pair $(X,x)$ of a type $X : \UU$ and an inhabitant $x:X$. We write $\UUpointed$ for the type of pointed types, 
 \begin{equation}
  \UUpointed \defeq \sm{X: \UU}X.
 \end{equation}
\end{defi}
\begin{defi}[Transitive Type]
 We say a type $X$ is \emph{transitive} and write $\istrans X$ if it satisfies 
 \begin{equation}
  \fapairs x y X \id[\UUpointed]{(X,x)}{(X,y)}.
 \end{equation}
\end{defi}

This is, of course, where univalence comes into play. It gives us the principle that a type $X$ is transitive if, and only if, for every pair $(x,y):X \times X$ there is an automorphism $e_{xy} : X \to X$ such that $\id{e_{xy}(x)}{y}$. 

We have the following examples of transitive types:
\begin{exa}
 Every type with decidable equality is transitive.
\end{exa}
This is because decidable equality on $X$ lets us define an endofunction on $X$ which swaps $x$ and $y$, and leaves everything else constant. 
Instances for this example include all contractible and, more generally, propositional types, but also our main candidate, the natural numbers $\N$.
\begin{exa}
 For any pointed type $X$ with elements $x_1, x_2 : X$, the identity type $\id[X]{x_1}{x_2}$ is transitive. In particular, the \emph{loop space} $\Omega^n(X)$ \cite[Definition 2.1.8]{HoTTbook} is transitive for any pointed type $X$.
\end{exa}
Here, it is enough to observe that, for $p_1, p_2 : \id[X]{x_1}{x_2}$, the function $\lam q q \ct \opp {p_1} \ct {p_2}$ is an equivalence with the required property.

As mentioned by Andrej Bauer in a discussion on this result~\cite{kraus:pseudoinverse}, we also have the following:
\begin{exa}
 Any group \cite[Definition 6.11.1]{HoTTbook} is a transitive type.
\end{exa}
As for equality types, the reason is that there is an inverse operation, such that the automorphism $\lam c c \ct \opp a \ct b$ maps $a$ to $b$.
\begin{exa}
 If $X$ is any type and $Y : X \to \UU$ is a family of transitive types, then $\prd{x:X}{Y(x)}$ is transitive.
\end{exa}
In particular, $\times$ and $\to$ preserve transitivity of types.

We are now ready to construct $\myst$:
Assume that we are given a type $X$. We can define a map
\begin{align}
 & f : X \to \UUpointed  \label{eq:f-for-myst} \\
 & f (x) \defeq (X,x). 
\end{align}
If we know a point $x_0 : X$, we may further define
\begin{align}
 & \overline f : \brck X \to \UUpointed \\
 & \overline f (z) \defeq (X, x_0).
\end{align}
If $X$ is transitive, we have 
\begin{equation}
 \fa x X \id{f(x)}{\overline f(\bproj x)}.
\end{equation}
By Theorem~\ref{thm:jdg-factor-nondep}, there is then a function 
\begin{equation}
 f' : \brck X \to \UUpointed
\end{equation}
such that, for any $x:X$, we have
\begin{equation}
 f'(\bproj x) \jdeq f(x) \jdeq (X,x).
\end{equation}
Let us define
\begin{align}
 & \myst_X : \prd{z:\brck X} \fst(f'(z)) \\
 & \myst_X \defeq \snd \circ f'. \label{eq:myst-definition}
\end{align}
Note that while the type of $\myst_X$ is \emph{not} simply $\brck X \to X$,
we have that, for any $x:X$, the type of $\myst_X(\bproj x)$ is judgmentally equal to $X$, and we have $\myst_X(\bproj x) \jdeq x$. 
This already proves the following:

\begin{thm} \label{thm:myst}
 Let $X$ be an inhabited transitive type. Then, there is a term $\myst_X$ such that the (dependent) composition
 \begin{align}
   & \myst_X \circ \bproj - : X \to X
\intertext{type-checks and is equal to the identity, where the proof}
  & p : \fa x X \id[X]{\myst_X(\bproj x)}{x} \\
  & p(x) \defeq \refl x
 \end{align}
 it trivial. \qed
\end{thm}

It is tempting to unfold the type expression $\prd{z:\brck X} \fst(f'(z))$ in order to better understand it. 
Unfortunately, this is not feasible as this plain type expression involves the whole proof term $f'$, which, in turn, includes the complete construction of Theorem~\ref{thm:jdg-factor-nondep}.
We want to emphasize again that, while we do have $\h_{\bproj x,\bproj y} : \id[\brck X]{\bproj x}{\bproj y}$ for any $x,y:X$, we cannot conclude $\id[X]{\myst_X(\bproj x)}{\myst_X(\bproj y)}$ 
as the expression $\mapfunc {\myst_X} (\h_{\bproj x,\bproj y})$ does not type-check.

Finally, we want to remark that the construction of $\myst$ does not need the full strength of Theorem~\ref{thm:jdg-factor-nondep}. The weaker version in which $\overline f : \brck X \to Y$ is replaced by a fixed $y_0 : Y$ is sufficient: in this case, $\overline f$ can be understood to be \emph{constant at $y_0$}. This leads to a simplification as the dependent function types in~\eqref{eq:jdgm-fact-g-type} and~\eqref{eq:jdgm-fact-lift-g-type} can be replaced by their codomains.

It may be helpful to see the whole definition of $\myst$ explicitly in this variant, which is also how it was explained originally by the first named author~\cite{kraus:pseudoinverse}:
We define
\begin{align}
 & \mathsf f : X \to \sm{A : \UUpointed} \; \id[\UUpointed]{A}{(X,x_0)} \\
 & \mathsf f (x) \defeq ((X,x) , \mathsf{transitive}_X(x,x_0)),
\end{align}
where $\mathsf{transitive}_X$ is the proof that $X$ is transitive. The function $f$ in~\eqref{eq:f-for-myst} is then simply the composition $\fst \circ \mathsf f$.
As the codomain of $\mathsf f$ is a singleton, it is contractible~(see~Definition~\ref{def:generalnotions}) and thereby propositional (let us write $h$ for the proof thereof). Hence, we get
\begin{align}
 & \mathsf{f'} : \brck{X} \to \sm{A : \UUpointed} \; \id[\UUpointed]{A}{(X,x_0)} \\
 & \mathsf{f'} \defeq \elim \left(\sm{A : \UUpointed} \; \id[\UUpointed]{A}{(X,x_0)}\right) \; h \; \mathsf f .
\end{align}
We could now define $\myst'_X$ to be 
\begin{align}
 &\myst'_X : \prd{\brck X} \fst \circ \fst \circ \mathsf{f'} \\
 &\myst'_X \defeq \snd \circ \fst \circ \mathsf{f'}
\end{align}
which has the same property as~\eqref{eq:myst-definition}, even though it is not judgmentally the same term.

\section{Conclusion and Open Problems} \label{sec10:open}

In this article, generalizations of Hedberg's Theorem have led us to an exploration of what we call \emph{weakly constant functions}. 
The attribute \emph{weakly} indicates that higher coherence conditions of such a constancy proof are missing. 
As a consequence, it is not possible to derive a function $\brck X \to Y$ from a weakly constant function $X \to Y$, but we have shown how to do this in several non-trivial special cases.
Most interesting is certainly the case of endofunctions. A weakly constant endofunction can always be factored through the propositional truncation of its domain. Further, for a given $X$, the type which says that every constant endofunction on $X$ has a fixed point is propositional, enabling us to use it as a notion of anonymous inhabitance $\populated X$, and we have argued that it lies strictly in between of $\neg\neg X$ and $\brck X$.

There are two questions for which we have not given an answer. 
The first is: Is weak propositional truncation definable in Martin-L\"of type theory? This is commonly believed to not be the case. However, the standard models do have propositional truncation, making it hard to find a concrete proof. Moreover, populatedness, a similar notion of anonymous existence, is definable.

Our second question is about the consequences of the assumption that weakly constant functions factor in general.  
By Shulman's result~\cite{shulman:wconst}, we know that this is inconsistent with the univalence axiom. 
Is is possible to strengthen this result further? 
In particular, does it imply UIP for all types? 
We leave these questions open.

\section*{Acknowledgements}
The first named author would like to thank Paolo Capriotti, Ambrus Kaposi, Nuo Li and especially
Christian Sattler for many fruitful discussions.
We are grateful to the anonymous referees for numerous helpful suggestions and remarks.
We also thank Nils Anders Danielsson for his careful reading of our draft and for pointing out several typos.

\bibliographystyle{plain}
\bibliography{hjReferences}

\end{document}